\documentclass[twocolumn,superscriptaddress]{revtex4-1} 

\usepackage{graphicx,epsfig} 
\usepackage{natbib} 
\usepackage[usenames,dvipsnames]{color} 
\usepackage{array}
\usepackage{amsmath} 
\usepackage{amssymb, bm}
\usepackage{soul} 
\usepackage{ifthen} 
\usepackage{upgreek}
\usepackage[ampersand]{easylist}
\usepackage{todonotes}
\usepackage{transparent}
\usepackage{xifthen}
\usepackage{xcolor}
\usepackage{verbatim}

\newcommand{\nn}{\nonumber}

\newcommand{\bra}[1]{\left\langle{#1}\right|}
\newcommand{\ket}[1]{\left|{#1}\right\rangle}

\def\l{\left}
\def\r{\right}
\def\be#1\ee{\begin{equation}#1\end{equation}}
\def\ba#1\ea{\begin{align}#1\end{align}}
\def\bg#1\eg{\begin{gather}#1\end{gather}}
\def\t{\text}

\newcommand{\abs}[1]{\lvert#1\rvert}

\newcolumntype{C}[1]{>{\centering\let\newline\\\arraybackslash\hspace{0pt}}m{#1}}

\begin{document}

\title{Tunable three-body coupler for superconducting flux qubits}

\author{D. Melanson}
\affiliation{Institute for Quantum Computing, and Department of Physics
and Astronomy, University
of Waterloo, Waterloo, ON, Canada N2L 3G1}

\author{A. J. Martinez}
\affiliation{Institute for Quantum Computing, and Department of Physics
	and Astronomy, University
	of Waterloo, Waterloo, ON, Canada N2L 3G1}

\author{S. Bedkihal}
\affiliation{Institute for Quantum Computing, and Department of Physics
	and Astronomy, University
	of Waterloo, Waterloo, ON, Canada N2L 3G1}

\author{A. Lupascu \footnotemark[1] \footnotetext[1]{Corresponding author: athree@uwaterloo.ca}}
\affiliation{Institute for Quantum Computing, and Department of Physics
and Astronomy, University
of Waterloo, Waterloo, ON, Canada N2L 3G1}
\affiliation{Waterloo Institute for Nanotechnology, University
	of Waterloo, Waterloo, ON, Canada N2L 3G1}

\date{ \today}

\begin{abstract}
The implementation of many-body interactions is relevant in various areas of quantum information. We present a superconducting device that implements a strong and tunable three-body interaction between superconducting quantum bits, with vanishing two-body interactions and robustness against noise and circuit parameter variations. These properties are confirmed by calculations based on the Born-Oppenheimer approximation, a two-level model for the coupling circuit, and numerical diagonalization. This circuit behaves as an ideal computational basis ZZZ coupler in a simulated three-qubit quantum annealing experiment. This work will be relevant for advanced quantum annealing protocols and future developments of high-order many-body interactions in quantum computers and simulators.
\end{abstract}

\maketitle

Many-body interactions can arise as mediated interactions from fundamental two-body couplings in various physical systems~\cite{Navr2003, EfimovTrimer2011}. In quantum information, many-body interactions are relevant for quantum error correction \cite{KITAEV20032,jiang_2017_noncommutingtwolocalhamiltonians}, complexity theory \cite{Kempe2005}, quantum thermodynamics \cite{Linden2010}, quantum chemistry \cite{AlanA2014}, and quantum simulations \cite{Quantsim2009,Porras2004}. The implementation of suitable high-order many-body interactions for quantum information comes with unique challenges, due to the fact that these interactions have to be of comparable strength with and controlled independently from lower-order interactions.

The implementation of many-body interactions has been considered in various physical systems for quantum information including ion traps~\cite{Bermudez2009}, atoms in optical lattices~\cite{Pachos2004, Semião2012}, and cold polar molecules~\cite{buchler_2007_threebodyinteractionscold}. In superconducting circuits, high-order effective interactions between qubits can be made strong, due to the fact that the underlying two-body interactions are strong. High-order interactions have been analyzed for several types of superconducting qubits and coupler circuits~\cite{Chen_2012, kafri_2017_tunableinductivecoupling, cho_2008_macroscopicmanyqubitinteractions, sametiSuperconductingQuantumSimulator2017, puri_2017_quantumannealingalltoalla}. Recently, there has been increased interest in many-qubit couplers for quantum annealers based on superconducting qubits. Proposals for engineered couplers for superconducting qubits that are suitable for quantum annealing include a three-body coupler circuit based on galvanic coupling \cite{ngcZZZAPS} and four-body coupler circuits based on a single-loop interferometer device  \cite{schondorf_2018_fourlocalinteractionssuperconducting} or a more complex circuit with a symmetric susceptibility used to cancel effectively lower-order interactions \cite{MITLL_ZZZZ_APS}. Other proposals rely on the use of ancilla qubits to reproduce the low-energy spectrum of many-body interactions, in the regime where the qubits have a Hamiltonian with negligible transverse terms in the basis of the interaction~\cite{chancellor_2017_circuitdesignmultibody}. However, designing circuits that implement many-body interactions and have other desirable features, including large tunable interaction strength, independence of the biasing conditions for the coupled qubits, cancellation of lower-order interactions, low design complexity, and robustness to noise and parameter variations, is a challenging problem.

In this Letter, we propose and analyze a superconducting quantum circuit used to implement a three-body interaction between three superconducting flux qubits. Flux-type qubits are promising candidates for both quantum annealing~\cite{harris_2010_experimentaldemonstrationrobust, weber_2017_coherentcoupledqubits} and gate-based quantum computing~\cite{yan_2016_fluxqubitrevisited}. The three-body coupler design that we propose is a relatively low-complexity circuit that combines strong tunable interactions of the order of 1 GHz, comparable to the qubit energy scales, effective cancellation of two-body terms, and robustness to noise and parameter variations.

\begin{figure}[h]
    \centering
    \includegraphics[width=\columnwidth]{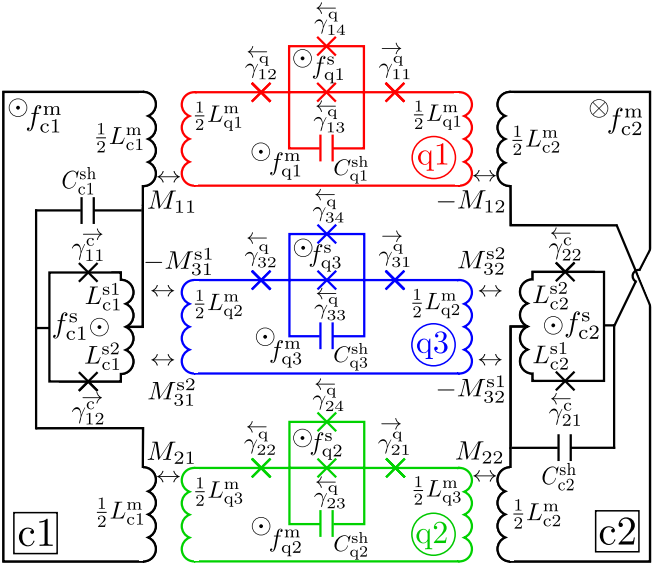}
    \caption{Circuit schematic of the coupler - qubits system. The two tunable rf-SQUID circuits forming the coupler (``c1'' and ``c2'') are coupled to the three tunable capacitively shunted flux qubits (``q1'', ``q2'', and ``q3'') by mutual inductances. Each loop of these circuits is subjected to a flux bias, as indicated. Josephson junctions are indicated by crosses (intrinsic junction capacitance not shown); arrows indicate the sign of the relative phase. Inductances of the secondary loops of the qubits are not shown. See text for additional details.}
    \label{fig:CouplerCircuit}
\end{figure}

Our proposed coupler device consists of two superconducting Josephson circuits, labeled ``c1'' and ``c2'' (see Fig. \ref{fig:CouplerCircuit}).  Each of these circuits has a main superconducting loop, interrupted by  two Josephson junctions which form a secondary loop. This circuit is referred to in other contexts as a tunable rf-SQUID~\cite{polettoTunableRfSQUID2009} or a compound Josephson junction rf-SQUID~\cite{harris_2009_compoundjosephsonjunctioncoupler}. The tunable rf-SQUID has the important property that the susceptibility, which is the relative change in current in the main loop due to a change in flux, can be controlled, in both sign and magnitude, by the magnetic fluxes applied to its two loops. Based on this property, a tunable rf-SQUID, operated either in the monostable~\cite{vandenbrink_2005_mediatedtunablecoupling,harris_2007_signmagnitudetunablecoupler} or quantum bistable~\cite{weber_2017_coherentcoupledqubits} regimes, can be used to mediate a tunable interaction between two flux qubits.

To achieve a strong three-body interaction, we employ the tunable rf-SQUID in the following way. We make the interaction between two qubits, each coupled to the rf-SQUID main loop, dependent on the state of a third qubit, coupled to the secondary loop, thereby generating an effective three-qubit interaction. However, in this coupler circuit based on only one tunable rf-SQUID, two-body interactions cannot be canceled or generally controlled independently from three-body interactions. To cancel the two-body interactions between qubits 1 and 2, we combine two rf-SQUIDs (circuits ``c1'' and ``c2'' in Fig.~\ref{fig:CouplerCircuit}) in such a way that qubit ``q1'' generates fluxes of equal value and opposite sign in the loops of couplers ``c1'' and ``c2'', by adding a twist in the main loop of ``c2''. To cancel the two-body interactions between qubits 1 and 3, and 2 and 3, we symmetrically couple ``q3'' to both branches of the coupler secondary loops and we bias ``c1''  and  ``c2'' at their main loop symmetry point, ensuring zero susceptibility  to  flux in  the  secondary  loop. Thus, the sum of the two-body interactions vanishes due to symmetry of the coupling and susceptibilities, with the susceptibilities controlled by the flux biasing conditions.

We next consider the full circuit representation of the coupler device and the qubits. In what follows, we assume that the qubits are capacitively shunted flux qubits (CSFQ), which are considered in recent efforts related to coherent quantum annealing \cite{weber_2017_coherentcoupledqubits}. However, the analysis applies straightforwardly to other flux qubit variants. The complete Hamiltonian of the qubits and coupler system is
\ba
\label{eq:Htotal}
H &= \sum_{i=1}^3 H_{\t{q}\,i} + \sum_{j=1}^2 H_{\t{c}\,j} + H_\t{int},
\ea
where $H_{\t{q}\,i}$ is the Hamiltonian of qubit $i$ ($i\in\{1, 2, 3\}$), $H_{\t{c}\,j}$ is the Hamiltonian of rf-SQUID $j$ ($j\in\{1, 2\}$), and $H_\t{int}$ is the interaction Hamiltonian. These Hamiltonians have the following form: 
\ba 
\label{eq:Hqubit}
H_{\text{q}i} &= \frac{1}{2 \hbar^2} \mathbf{p}_{\t{q}i}^\t{T}\, \mathbf{E}_{\t{q}i}^\t{C}\, \mathbf{p}_{\t{q}i} -\sum_{k=1}^4\! E_{\t{q}ik}^{\text{J}} \cos{\gamma_{ik}^{\t{q}}} + \frac{1}{2} \boldsymbol{\varphi}_{\t{q}i}^\t{T}\, \mathbf{E}_{\t{q}i}^\t{L}\, \boldsymbol{\varphi}_{\t{q}i},
\ea
\ba
\label{eq:Hcoupler}
H_{\text{c}j} &= \frac{1}{2 \hbar^2} \mathbf{p}_{\t{c}j}^\t{T}\, \mathbf{E}_{\t{c}j}^\t{C}\, \mathbf{p}_{\t{c}j} - \sum_{l=1}^2 E_{\t{c}jl}^{\text{J}} \cos{\gamma_{jl}^{\t{c}}} + \frac{1}{2} \boldsymbol{\varphi}_{\t{c}j}^\t{T}\, \mathbf{E}_{\t{c}j}^\t{L}\, \boldsymbol{\varphi}_{\t{c}j}
\ea
and
\be
\label{eq:Hint}
H_\text{int} = \frac{\phi_0^2}{2} \boldsymbol{\varphi}^\t{T}\, \l( \mathbf{L}^{-1}-\mathbf{L}_0^{-1} \r) \boldsymbol{\varphi}.
\ee
In Eqs.~(\ref{eq:Hqubit})-(\ref{eq:Hint}), the independent degrees of freedom are the phases $\gamma^\t{q}_{i\,k}$ ($\gamma^\t{c}_{j\,l}$) across junction $k$ ($l$) and the phase $\varphi^\t{m}_{\t{q}\,i}$ ($\varphi^\t{m}_{\t{c}\,j}$) across the main loop inductance of the qubit (coupler) $i$ ($j$) for $i \in \l\{ 1,2,3 \r\}$ ($j \in \l\{ 1,2 \r\}$) and $ k \in \l\{ \overline{1,4} \r\}$ ($ l \in \l\{ 1,2 \r\}$). The $p^\alpha_{i\,k}$ are the conjugate momenta to each $\gamma^\alpha_{i\,k}$, and similarly $p^\t{m}_{\alpha\,i}$ conjugate momenta to each $\varphi^\t{m}_{\alpha\,i}$. The vectors of inductance phases are $\boldsymbol{\varphi}^\t{T}_{\alpha\,i} = \l( \varphi_{\alpha i}^{\t{m}}, \varphi_{\alpha i}^{\t{s}1}, \varphi_{\alpha i}^{\t{s}2} \r)$ for $\alpha i \in \l\{ \t{q}1, \t{q}2, \t{q}3,  \t{c}1, \t{c}2 \r\}$ and $\boldsymbol{\varphi}^\t{T} = \l( \boldsymbol{\varphi}^\t{T}_{\t{q}\,1}, \boldsymbol{\varphi}^\t{T}_{\t{q}\,2}, \boldsymbol{\varphi}^\t{T}_{\t{q}\,3}, \boldsymbol{\varphi}^\t{T}_{\t{c}\,1}, \boldsymbol{\varphi}^\t{T}_{\t{q}\,2} \r)$, with $ \varphi^\t{s1}_{\t{q}\,i} = \gamma^\t{q}_{i\,1} - \gamma^\t{q}_{i\,2} - \gamma^\t{q}_{i\,4} - \varphi^\t{m}_{\t{q}\,i} - 2 \pi f_{\t{q}\,i}^{\t{m}} - \pi f_{\t{q}\,i}^{\t{s}} $, $ \varphi^\t{s2}_{\t{q}\,i} = \gamma^\t{q}_{i\,1} - \gamma^\t{q}_{i\,2} - \gamma^\t{q}_{i\,3} - \varphi^\t{m}_{\t{q}\,i} - 2 \pi f_{\t{q}\,i}^{\t{m}} + \pi f_{\t{q}\,i}^{\t{s}} $, $ \varphi^\t{s1}_{\t{c}\,j} = -\gamma^\t{c}_{j\,1} - \varphi^\t{m}_{\t{c}\,j} - 2 \pi f_{\t{c}\,j}^{\t{m}} + \pi f_{\t{c}\,j}^{\t{s}} $ and $ \varphi^\t{s2}_{\t{c}\,j} = -\gamma^\t{c}_{j\,2} - \varphi^\t{m}_{\t{c}\,j} - 2 \pi f_{\t{c}\,j}^{\t{m}} - \pi f_{\t{c}\,j}^{\t{s}}  $. The vectors of conjugate momenta are $ \mathbf{p}^\t{T}_{\t{q}\,i}=\l( p^\t{q}_{i\,1},\, p^\t{q}_{i\,2},\, p^\t{q}_{i\,3},\, p^\t{q}_{i\,4}, p^\t{m}_{\t{q}\,i} \r)$, $ \mathbf{p}^\t{T}_{\t{c}\,j}=\l( p^\t{c}_{j\,1},\, p^\t{c}_{j\,2}, p^\t{m}_{\t{c}\,j} \r)$. The kinetic energy of each device is $\mathbf{E}_{\alpha\,i}^\t{C} = 4 e^2\, \mathbf{C}_{\alpha\,i}^{-1}$, with $e$ the electron charge and $\mathbf{C}_{\alpha\,i}$ the capacitance matrix for device $\alpha\,i$, composed of the junction capacitances $C^\alpha_{ik}$ and the shut capacitance $C^\t{sh}_{\alpha i}$. The Josephson energy for junction $k$ of device $\alpha \,i$ is $E_{\alpha\,i\,k}^\t{J} = \phi_0\, I^\t{J}_{\alpha\,i\,k}$, where $\phi_0 = \hbar / 2 e$, with $\hbar$ the Plank constant, is the reduced magnetic flux quantum and $I^\t{J}_{\alpha\,i\,k}$ the critical current of the junction. The inductive energy of each device is $\mathbf{E}_{\alpha\,i}^\t{L} = \phi_0^2\, \mathbf{L}_{\alpha\,i}^{-1}$, with $\mathbf{L}_{\alpha\,i}$ the inductance matrix for device $\alpha\,i$, composed of the main loop inductance $L_{\alpha\,i}^{\t{m}}$ and of the secondary loop inductance, which is divided into two branches $L_{\alpha\,i}^{\t{s}1}$ and $L_{\alpha\,i}^{\t{s}2}$. Similarly, $\mathbf{L}$ is the inductance matrix of the coupled system, the diagonal elements of which are the loop inductances $L_{\alpha\,i}^\beta$ for $\beta \in \l\{\t{m,s1,s2}\r\}$ of all devices $\alpha i \in \l\{ \t{q}1, \t{q}2, \t{q}3,  \t{c}1, \t{c}2 \r\}$ and the off-diagonal elements of which are the mutual inductances $M_{i,j}$ for $i,j \in \l\{1,2\r\}$ and $M_{i,3}^\beta$ for $1 \in \l\{1,2,3\r\}$ and $\beta \in \l\{\t{s1,s2}\r\}$ between devices, and $\mathbf{L}_0$ is the diagonal inductance matrix of the uncoupled system (equivalent to $\mathbf{L}$ with vanishing mutual inductance). Additionally, external bias magnetic fluxes $f_{\alpha\, i}^{\beta}$ thread each loop $\beta \in \l\{ \t{m,s} \r\}$, of each device with $\alpha i \in \l\{ \t{q}1, \t{q}2, \t{q}3,  \t{c}1, \t{c}2 \r\}$, and we assign these to the three inductances $L_{\alpha\,i}^\beta$ for $\beta \in \l\{\t{m,s1,s2}\r\}$ of the devices. Finally, in what follows, we take identical qubits and identical rf-SQUIDs with the following values of the circuit parameters: for qubits $i$ ($i \in \l\{1,2,3\r\}$), $I^\t{J}_{\t{q}\,i\,k} = 221.0 ~\t{nA}$ and $C^\t{q}_{ik} = 4.5 ~\t{fF}$ for $k\in\l\{1,2\r\}$, $I^\t{J}_{\t{q}\,i\,k} = 102.0 ~\t{nA}$ and $C^\t{q}_{ik} = 2.0 ~\t{fF}$ for $k\in\l\{3,4\r\}$, $C^\t{sh}_{\t{q} i} = 40.0 ~\t{fF}$, $L_{\t{q}\,i}^{\t{m}} = 250.0 ~\t{pH}$, $L_{\t{q}\,i}^{\t{s}1} = L_{\t{q}\,i}^{\t{s}2} = 10.0 ~\t{pH}$, and for coupler $j$ ($j \in \l\{1,2\r\}$)
$I^\t{J}_{\t{c}\,j\,l} = 600.0 ~\t{nA}$ and $C^\t{c}_{jl} = 12.0 ~\t{fF}$ for $l \in \l\{1,2\r\}$, $C^\t{sh}_{\t{c} j} = 10.0 ~\t{fF}$, $L_{\t{c}\,i}^{\t{m}} = 550.0 ~\t{pH}$, $L_{\t{c}\,i}^{\t{s}1} = L_{\t{c}\,i}^{\t{s}2} = 85.0 ~\t{pH}$, and mutual inductances of $M_{i\,j} = - 50.0$ pH for $i,j \in \l\{1,2\r\}$ and $M_{3\,j}^\t{s1} =  M_{3\,j}^\t{s2} = - 25.0$ pH for $j \in \l\{1,2\r\}$. 

We first discuss the properties of the coupler using the Born-Oppenheimer approximation, which validates the intuitive picture for the three-body coupler based on rf-SQUID susceptibility. Since the tunable rf-SQUID circuits are designed to have a significantly larger energy gap between the ground and excited states (typical gap of 15 GHz) than the qubits (typical gap of 4 GHz), it is reasonable to apply the Born-Oppenheimer approximation \cite{Massey_1949,vandenbrink_2005_mediatedtunablecoupling,kafri_2017_tunableinductivecoupling}. This is done by replacing $H$ with the effective qubit Hamiltonian
\be
\label{eq:HeffBO}
H_{\t{eff}}= \sum_{i=1}^3 H_{\text{q}\,i} +
\widetilde{E}_{\text{c}}^{(0)} \l( \l\{ f_{\t{c}\, j}^\beta + \delta f_{\t{q}\,j}^\beta \r\}_{ \substack{ \beta \in \{\t{m}, \t{s}\} \\ j \in \{1,2\} } } \r),
\ee
where $\widetilde{E}_{\text{c}}^{(0)}$ is the quantum ground state energy of $\widetilde{H}_{\text{c}}=\sum_{j=1}^2 H_{\text{c}\,j}+H_{\text{int}}$. Here, $\widetilde{E}_{\text{c}}^{(0)}$ is an operator, dependent on the effective qubit flux shifts $\delta f_{\t{q}\,j}^\t{m} = \sum_{i = 1}^2 I_{\t{q}i}^\t{m}\, M_{i\,j}$ and $\delta f_{\t{q}\,j}^\t{s} = I_{\t{q}3}^\t{m} \l( M_{3\,j}^\t{s1} + M_{3\,j}^\t{s2} \r)$, with $I_{\t{q}i}^\t{m} =  \phi_0\, \varphi_{\t{q}\,i}^\t{m} / L_{\t{q}\,i}^\t{m}$ the operator for the circulating current in the main loop of qubit $i$. 

\begin{figure}[h]
	\centering
	\includegraphics[width=\columnwidth]{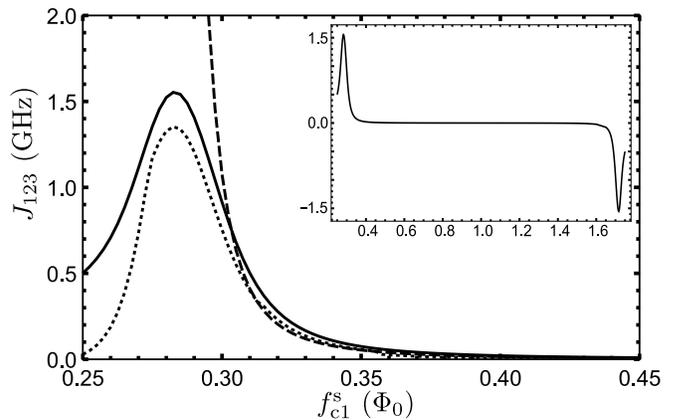}
	\caption{Three-body coupling strength extracted by Born-Oppenheimer method in the circuit model (solid line), perturbation theory in the spin model (dashed line), and numerical calculations of the full circuit (dotted line), versus the $f_{\t{c}\, 1}^{s}$. The flux biases are $f_{\t{c}\,1}^{\t{m}}=f_{\t{c}\,2}^{\t{m}}=1/2$, $f_{\t{c}\,1}^{\t{s}}=-f_{\t{c}\,2}^{\t{s}}$, $f_{\t{q}\,i}^{\t{m}}=1/2$ and $f_{\t{q}\,i}^{\t{s}}=0$ for $i \in \l\{ 1,2,3 \r\}$. The inset shows the coupling strength calculated using Born-Oppenheimer approximation over a wide flux range.}
	\label{fig:Kcompare}
\end{figure}

The analysis of the Born-Oppenheimer approximation is simplified if we project $\widetilde{E}_{\text{c}}^{(0)}$ onto the subspace formed by the two lowest energy states of each qubit, which is appropriate because of the large anharmonicity in the spectrum of the CSFQ \cite{yan_2016_fluxqubitrevisited,weber_2017_coherentcoupledqubits}. The state of each qubit $i$ is represented by the binary current variable $I^\t{p,m}_{i}\, s_{i}$, where $s_i= \pm 1 $ and $I^\t{p,m}_{\t{q}i}$ is the persistent current in the main loop of qubit $i$. The persistent current is defined as the average of the magnitudes of the eigenvalues of the current operator represented in the basis of the two lowest energy states of each qubit. In terms of the binary qubit parameters, the energy is written as
\ba
\label{eq:E0expansion}
\widetilde{E}_{\text{c}}^{(0)} \l( s_1,s_2,s_3 \r) &= \widetilde{A}_0 + \sum_{i=1}^3 \tilde{h}_{i}\,s_{i} + \sum_{\substack{i,\,j=1\\ i\neq j}}^3 \widetilde{J}_{ij}\, s_{i} s_{j} \nn\\
& \quad + \widetilde{J}_{123}\, s_1s_2s_3,
\ea
where $\widetilde{A}_0$ is an energy offset, $\tilde{h}_{i}$ are single qubit biases arising from screening currents in the coupler, $\widetilde{J}_{ij}$ are two-body interactions, and $\widetilde{J}_{123}$ is the 3-local coupling strength. We determine these coefficients by solving the linear system of equations given by Eq.~\ref{eq:E0expansion} for the 8 different values of qubit variables [see supplementary information (SI)]. To calculate $\widetilde{E}_{\text{c}}^{(0)} \l( s_1,s_2,s_3 \r)$ for each triplet of $s_i = \pm 1$, we numerically diagonalize $\widetilde{H}_{\text{c}}$, using a representation in terms of the harmonic oscillator states for the quadratic part of the Hamiltonian. The three-body coupling strength $\widetilde{J}_{123}$ is shown in Fig.~\ref{fig:Kcompare} versus $f_{\t{c}\,1}^{\t{s}}$, with $f_{\t{c}\,1}^{\t{s}}=-f_{\t{c}\,2}^{\t{s}}$ and all the qubits biased as $f_{\t{q}\,i}^{\t{m}}=1/2$ and $f_{\t{q}\,i}^{\t{s}}=0$ for $i \in \l\{ 1,2,3 \r\}$. With this biasing condition, all two-body terms cancel, in agreement with the qualitative picture presented above.

We discuss next a useful model for the coupler, in which each of the two tunable rf-SQUIDs is modeled as a two-level system. The two-level approximation is reasonable due to the large energy gap between the first and second excited states of each rf-SQUID. This simple model, in which both qubits and the coupling circuits are treated as spins, allows to obtain perturbative analytical expressions for the effective mediated interactions, giving additional insight into the valid parameter range of the coupler and on the required conditions for cancellation of two-body interactions. 

In this spin model, the full Hamiltonian is 
\begin{equation}
\label{eq:HSpin}
\bar{H} = \bar{H}_{\t{Q}} + \bar{H}_{\t{C}} + \bar{H}_{\text{Int}},
\end{equation}
where
\ba
\bar{H}_{\t{Q}} &= \sum_{i=1}^{3} \left[ \frac{\bar{\Delta}_{\t{Q}\,i}}{2}\, \sigma_{\t{Q}\,i}^{x} + \frac{\bar{\epsilon}_{\t{Q}\,i}}{2}\, \sigma_{\t{Q}\,i}^{z} \right],\\
\bar{H}_{\t{C}} &= \sum_{j=1}^{2} \left[ \frac{\bar{\Delta}_{\t{C}\,j}}{2}\, \sigma_{\t{C}\,j}^{x} + \frac{\bar{\epsilon}_{\t{C}\,j}}{2}\, \sigma_{\t{C}\,j}^{z} \right]
\ea
and
\be
\bar{H}_{\text{Int}} = \sum_{i=1}^{2} \sum_{j=1}^{2} \bar{J}_{\t{Q}i,\t{C}j}\, \sigma_{\t{Q}\,i}^{z}\, \sigma_{\t{C}\,j}^{z} + \sum_{j=1}^{2} \bar{J}_{\t{Q}3,\t{C}j}\, \sigma_{\t{Q}\,3}^{z}\, \sigma_{\t{C}\,j}^{x}
\ee
are the Hamiltonians for the qubits, coupler, and the interaction between them, respectively. To differentiate the spin model parameters, we use an over-bar and we use capital letters Q and C to denote qubits and coupling circuits. Here, $\sigma_{\alpha\,i}^{x}$ and $\sigma_{\alpha\,i}^{z}$ are the Pauli matrices, $\bar{\Delta}_{\alpha\,i}$ and $\bar{\epsilon}_{\alpha\,i}$ are the tunneling and bias of device $\alpha\,i$. The coupling between the qubit $i$ and coupler $j$ is, for $i\,j \in \l\{1,2\r\}$ $\bar{J}_{\t{Q}i,\t{C}j} = M_{i\,j}\,I^\t{p,m}_{\t{q}\,i}\,I^\t{p,m}_{\t{c}\,j}$ and $\bar{J}_{\t{Q}3,\t{C}j} = \l( M_{3\,j}^\t{s1} + M_{3\,j}^\t{s2} \r)\,I^\t{p,m}_{\t{q}\,i}\,I^\t{p,s}_{\t{c}\,j}$, where $I^\t{p,m}_{\t{c}\,j}$ ($I^\t{p,s}_{\t{c}\,j}$) is the persistent current in the coupler main (secondary) loop. We determine the relevant model parameters $\bar{\epsilon}_{\alpha i}$ and $\bar{\Delta}_{\alpha i}$ for $\alpha \in \l\{ \t{Q}, \t{C} \r\}$, based on the properties of the lowest two energy eigenstates of each qubit/coupler unit (see SI).
We assume weak direct qubit-coupler interaction, i.e. $\bar{J}_{\t{Q}i,\t{C}j} / \omega_{0,1} \ll 1$, where $\omega_{0,1}$ is the coupler excitation energy. We then integrate out the coupler degrees of freedom, and derive an effective Hamiltonian in the subspace of slow degrees of freedom of the qubits. The effective interaction is obtained by projecting the full time evolution operator (in the interaction picture) onto the ground state of the coupler. Following the perturbative procedure in Ref.~\cite{Hutter_2006}, we obtain the following effective qubit Hamiltonian:
\ba
\label{eq:HeffSpin}
\bar{{H}}_{\text{eff}}& =
\sum_{i} \frac{1}{2} \left[ \left( \bar{\Delta}_{\t{Q}\,i} + \bar{\Delta}'_{\t{Q}\,i} \right) \sigma_{\t{Q}\,i}^{x} + \left( \bar{\epsilon}_{\t{Q}\,i} + \bar{\epsilon}'_{\t{Q}\,i} \right) \sigma_{\t{Q}\,i}^{z} \right]\nn \\
& \quad + \bar{J}_{12}\, \sigma_{\t{Q}\,1}^{z}\, \sigma_{\t{Q}\,2}^{z} + \bar{J}_{13}\, \sigma_{\t{Q}\,1}^{z}\, \sigma_{\t{Q}\,3}^{z} + \bar{J}_{23}\, \sigma_{\t{Q}\,2}^{z}\, \sigma_{\t{Q}\,3}^{z}\nn \\
& \qquad + \bar{J}_{123}\, \sigma_{\t{Q}\,1}^{z}\, \sigma_{\t{Q}\,2}^{z}\, \sigma_{\t{Q}\,3}^{z}.
\ea
In Eq.~(\ref{eq:HeffSpin}),
$\bar{\Delta}'_{\t{Q}\,i}$, and $\bar{\epsilon}'_{\t{Q}\,i}$ represent the qubit detuning and bias shifts induced by the coupler (see SI), $\bar{J}_{12} = - \sum_{n} \bar{V}_{0,n} \left( \bar{J}_{\t{Q}1,\t{C}1}\, \bar{J}_{\t{Q}2,\t{C}2} + \bar{J}_{\t{Q}2,\t{C}1}\, \bar{J}_{\t{Q}1,\t{C}2} \right) /  \omega_{n,0}$, with similar expressions for $\bar{J}_{13}$ and $\bar{J}_{23}$ (see SI). Here, $\bar{V}_{0,n} = \bra{0_\t{c}} \sigma_{\t{C}1}^{z} \ket{n_\t{c}} \bra{n_\t{c}} \sigma_{\t{C}2}^{z} \ket{0_\t{c}}$ are the matrix elements of the coupler operators corresponding to the ground and $n^\t{th}$ level of the coupler. Finally, in a compact form, $\bar{J}_{123}= \sum_{n} \sum_{m} \bra{0_\t{c}} \bar{H}_{\t{int}}\, \mathcal{P}_n\, \bar{H}_{\t{int}}\, \mathcal{P}_m\, \bar{H}_{\t{int}}\, \ket{0_\t{c}} / \l( \omega_{0,n}\, \omega_{0,m} \r)$, where $\mathcal{P}_a = \ket{a_\t{c}} \bra{a_\t{c}}$, for $a \in \l\{ n,m \r\}$ are projection operators in the coupler subspace (see SI). Figure \ref{fig:Kcompare} shows $\bar{J}_{123}$ versus $f_{\t{c}\,j}^{\t{s}}$.
The spin model predicts the cancellation of the two-body term $\bar{J}_{12}$, and similarly of the other two-body interactions, for correct choice of bias conditions. These cancellation conditions are in agreement with our intuitive picture based on symmetries of coupling and susceptibilities.

To validate the above approximate treatments, we numerically compute the eigenstates of the complete Hamiltonian Eq.~\eqref{eq:Htotal}. We represent each non-periodic (periodic) degree of freedom of the Hamiltonian in a basis formed of harmonic oscillator (Fourier) states. Due to the complexity of the complete circuit, we proceed hierarchically by first diagonalizing each device separately. Then, keeping only the low-energy eigenstates of each device, we introduce the coupling between all devices and diagonalize the combined system in this low-energy basis to calculate the energy spectrum \cite{JJSim}. The two-body and three-body interactions are determined based on energy gaps at anti-crossings in the numerically calculated spectrum (see SI for details).

The three-body interaction strength $\bar{J}_{123}$ calculated based on the spin model is in excellent agreement with numerical results for the full circuit model for $\abs{f_{\t{c}\,j}^{\t{s}}} > 0.31$, as shown in  Fig.~\ref{fig:Kcompare}. The disagreement for small values of $\abs{f_{\t{c}\,j}^{\t{s}}}$ is due to the increase in the ratio $\bar{J}_{\t{Q}i,\t{C}j} / \omega_{0,1}$, arising from both a decreasing energy gap of the rf-SQUIDs and the increase in the qubit-rf-SQUID coupling. We also find good qualitative agreement between the Born-Oppenheimer method and the numerical calculations. However, the pronounced quantitative disagreement for $\abs{f_{\t{c}\,j}^{\t{s}}} < 0.31$ between the two methods can be attributed to the breakdown of the Born-Oppenheimer approximation in the region of decreasing energy gap of the rf-SQUID \cite{kafri_2017_tunableinductivecoupling}. 

An important feature for a superconducting coupler is robustness to low-frequency flux noise \cite{yoshihara_2006_decoherencefluxqubits}. We assume Gaussian flux noise with the power spectral density $A_{\t{c}\,i}^\beta / \abs{\omega}^\gamma$  \cite{koch_2007_modelfluxnoise}, where $\omega$ is the angular frequency. We use parameters $A_{\t{c}\,j}^\t{m} = 14.6 ~\mu\Phi_0 / \sqrt{\t{Hz}}$,  $A_{\t{c}\,j}^{\t{s}i} = 4.1~\mu\Phi_0 / \sqrt{\t{Hz}}$ for $i \in \l\{1,2\r\}$, and $\gamma=0.9$, based on experiments \cite{bialczak_2007_fluxnoisejosephson,yan_2016_fluxqubitrevisited,weber_2017_coherentcoupledqubits} (see SI). We treat the inductances of the SQUID as three, uncorrelated, contributions to the flux noise. At the bias point where we get large three-body coupling and a cancellation of all two-body couplings ($f_{\t{c}\,1}^{\t{m}}=f_{\t{c}\,2}^{\t{m}}=0.5$ and $f_{\t{c}\,1}^{\t{s}}=-f_{\t{c}\,2}^{\t{s}}=0.3$), we observe a standard deviation in $\widetilde{J}_{123}$ of at most 0.003 GHz around the nominal value of 1 GHz as well as a standard deviation of at most 0.015 GHz around the nominal cancellation point for $\widetilde{J}_{12}$, $\widetilde{J}_{13}$ and $\widetilde{J}_{23}$ (see SI for all coefficients). The spread in single qubit energy bias terms $\tilde{h}_i$ induced by the coupler circulating currents is comparable with fluctuations induced by flux noise intrinsic to the qubit loops. As a result, we do not expect the coupler to induce significant dephasing of the qubit. These small spreads in parameters ensures that the correct solution is found at the end of the computation \cite{perdomo-ortiz_2016_determinationcorrectionpersistent}.

Besides robustness to flux noise, the coupler should be functional when variations in fabrication parameters are taken into account. To validate robustness to errors in junction critical currents and self and mutual inductances, we extracted the effective qubit Hamiltonian over a normal distribution of parameter variations, with standard deviation of 3\% of the nominal values. We find that inductance variations in the secondary loop have a negligible effect on the two- and three- body coupling strengths. However, main loop self inductance and junction asymmetry significantly affects the values of the $\widetilde{J}_{ij}$, but cancellation conditions for two-body terms, to within 10 MHz, can be recovered by suitable compensations of the bias fluxes (see SI). 
\begin{figure}[h]
    \centering
    \includegraphics[width=\columnwidth]{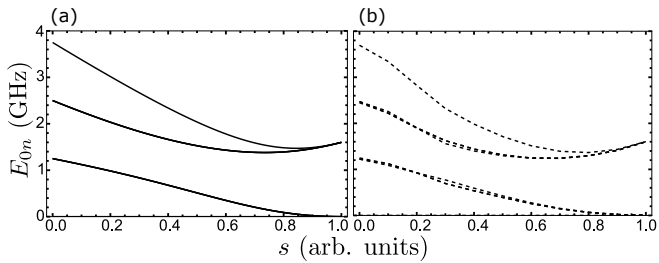}
	\caption{Low-energy spectrum versus annealing parameter $s$. (a) Spectrum of the 8 levels of an ideal 3-spin Hamiltonian implementing Eq.~\eqref{eq:Hanneal}. (b) Spectrum of the lowest 8 qubit-like levels of the circuit Hamiltonian \eqref{eq:Htotal} when biased to implement a linear annealing schedule as in \eqref{eq:Hanneal}.}
	\label{fig:AnnealinSpect}
\end{figure}

When implementing this coupler in a quantum annealing context, an important metric is the extent to which the coupler emulates the spectrum of an ideal spin Hamiltonian. In quantum annealing~\cite{albash_2018_adiabaticquantumcomputation}, one seeks to prepare the ground state of a Hamiltonian that encodes the solution to a computational problem. The solution is found by initializing the computation in the ground state of a trivial Hamiltonian, followed by adiabatically transforming this Hamiltonian into the Hamiltonian of interest. We consider an annealing schedule with an initial Hamiltonian given by the standard transverse field Hamiltonian and a final Hamiltonian given by a three-body interaction:
\ba
\label{eq:Hanneal}
H_{\text{anneal}}(s) = \l( 1-s \r) \sum_i^3 \frac{\Delta_i}{2}\, \sigma^x_i + s\, J_{123}\, \sigma^z_1 \sigma^z_2 \sigma^z_3,
\ea
where $s$ is the annealing parameter, changing from 0 to 1. Figure~\ref{fig:AnnealinSpect} shows the energy levels of the Hamiltonian Eq.~\eqref{eq:Hanneal} and of the complete circuit, including the three body coupler, where we have used $\Delta_i = 1.22$ GHz and $J_{123} = 0.8$ GHz (see SI). We find excellent agreement between the energy spectra in the two cases, and in particular the physical system correctly preserves the degeneracy of the ideal spin Hamiltonian. We also have that the lowest 8 qubit-like energy levels of the complete circuit are well separated from the higher levels (gap of approximately 8 GHz).

We proposed and analyzed a three-local coupler for superconducting flux qubits. This coupler is sign and magnitude tunable, allows for cancellation of two-body interactions, and is robust against noise and fabrication parameter variations. A system formed of three qubits and the coupler has the correct spectral properties of an ideal spin Hamiltonian. The analysis presented in this work will be relevant in designing the next generation of quantum annealing hardware which will have capabilities in increased embedding efficiency for problems that include three-body terms, in quantum annealing error correction and quantum simulations. 


We acknowledge useful discussions with M. A. Yurtalan, Y. Tang, G. Consani, P. W. Warburton, M. Sch\"ondorf, F. K. Wilhelm, S. Novikov, K. Zick, D. Ferguson, S. Weber, A. J. Kerman, M. Khezri, D. Lidar. This material is based upon work supported by the Intelligence Advanced Research Projects Activity (IARPA) and the Army Research Office (ARO) under Contract No. W911NF-17-C-0050. Any opinions, findings and conclusions or recommendations expressed in this material are those of the author(s) and do not necessarily reflect the views of the Intelligence Advanced Research Projects Activity (IARPA) and the Army Research Office (ARO). 





%

\pagebreak
\begin{center}
\textbf{\large Supplementary Information: Tunable three-body coupler for superconducting flux qubits}
\end{center}

\setcounter{equation}{0}
\setcounter{figure}{0}
\setcounter{table}{0}
\setcounter{page}{1}
\makeatletter
\renewcommand{\theequation}{S\arabic{equation}}
\renewcommand{\thefigure}{S\arabic{figure}}
\renewcommand{\bibnumfmt}[1]{[S#1]}
\renewcommand{\citenumfont}[1]{S#1}

\section{Born-Oppenheimer Inversion Method to Extract Coupling Strength}
The interaction coefficients in Eq.~(\ref{eq:E0expansion}) of the main text, $\tilde{A}_0$, $\tilde{h}_{i}$, $\tilde{J}_{ij}$, and $\tilde{J}_{123}$ are found by numerically calculating the ground state energy of the coupler Hamiltonian for all eight configurations of the binary qubit variables. These eight energies are related to the coupling coefficients as
\ba
\label{eq:BOinversion}
\begin{bmatrix}
	\widetilde{E}_{\text{c}}^{(0)}(-1,-1,-1)\\
	\widetilde{E}_{\text{c}}^{(0)}(-1,-1,~~1)\\
	\widetilde{E}_{\text{c}}^{(0)}(-1,~~1,-1)\\
	\widetilde{E}_{\text{c}}^{(0)}(~~1,-1,-1)\\
	\widetilde{E}_{\text{c}}^{(0)}(-1,~~1,~~1)\\
	\widetilde{E}_{\text{c}}^{(0)}(~~1,-1,~~1)\\
	\widetilde{E}_{\text{c}}^{(0)}(~~1,~~1,-1)\\
	\widetilde{E}_{\text{c}}^{(0)}(~~1,~~1,~~1)
\end{bmatrix}
=
\mathcal{S}
\begin{bmatrix}
	\tilde{A}_0\\\tilde{h}_1\\\tilde{h}_2\\\tilde{h}_3\\\tilde{J}_{1\,2}\\\tilde{J}_{1\,3}\\\tilde{J}_{2\,3}\\\tilde{J}_{123}
\end{bmatrix},
\ea
with
\ba
\mathcal{S}=
\begin{bmatrix}
	& 1 & -1 & -1 & -1 & ~~1 & ~~1 & ~~1 & -1 \\
	& 1 & -1 & -1 & ~~1 & ~~1 & -1 & -1 & ~~1 \\
	& 1 & -1 & ~~1 & -1 & -1 & ~~1 & -1 & ~~1 \\
	& 1 & ~~1 & -1 & -1 & -1 & -1 & ~~1 & ~~1 \\
	& 1 & -1 & ~~1 & ~~1 & -1 & -1 & ~~1 & -1 \\
	& 1 & ~~1 & -1 & ~~1 & -1 & ~~1 & -1 & -1 \\
	& 1 & ~~1 & ~~1 & -1 & ~~1 & -1 & -1 & -1 \\
	& 1 & ~~1 & ~~1 & ~~1 & ~~1 & ~~1 & ~~1 & ~~1
	\nn
\end{bmatrix}.
\ea
Coefficients $\tilde{A}_0$, $\tilde{h}_{i}$, $\tilde{J}_{ij}$, and $\tilde{J}_{123}$ are found by inverting the system of equations~(\ref{eq:BOinversion}).

\section{Extracting coupling strength from avoided level crossing}
\begin{figure}[!]
	\centering
	\includegraphics[width=\columnwidth]{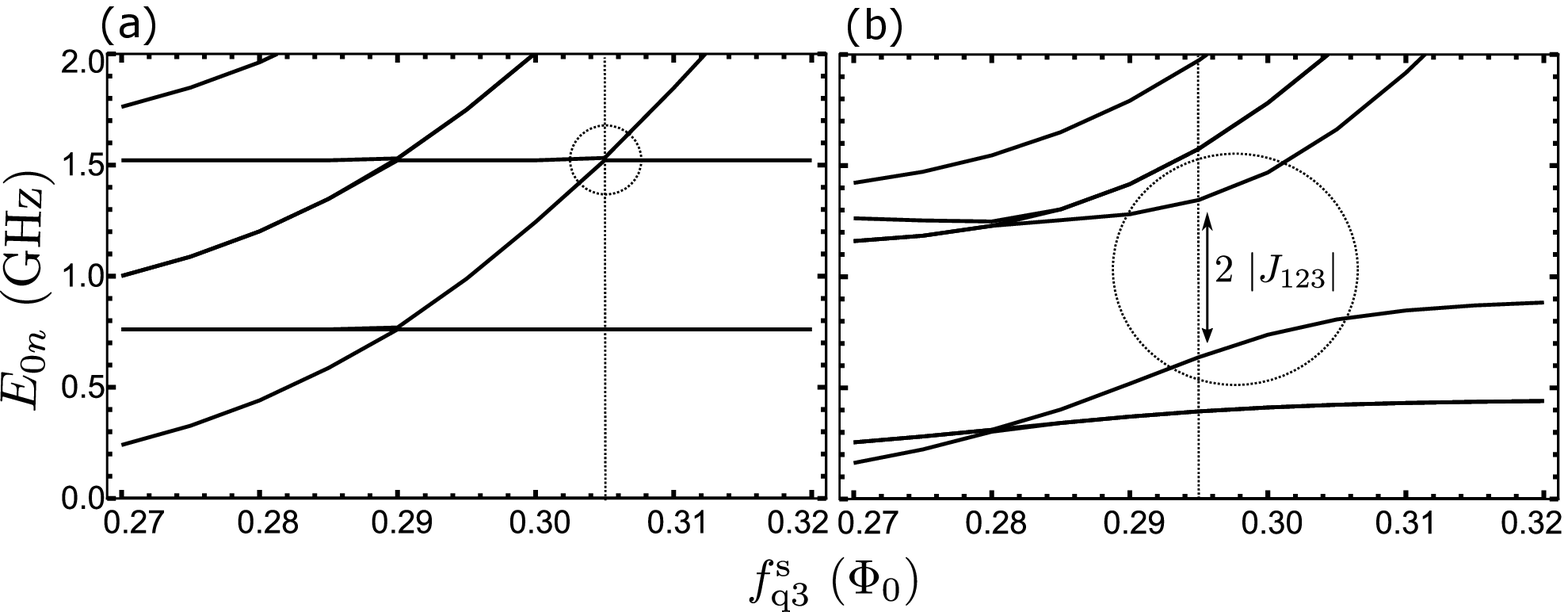}
	\caption{Energy spectrum of the complete circuit Hamiltonian, relative to the ground state energy. (a) When the three-body coupling is turned on, an avoided level crossing between level 4 and 5 corresponding to $2\, \abs{J_{123}}$ appears around $\Delta_{\t{Q}3} \approx \Delta_{\t{Q}1} + \Delta_{\t{Q}2}$ (dotted line and circle). (b) When the three-body coupling is turned off, levels 4 and 5 simply cross each other around $\Delta_{\t{Q}3} \approx \Delta_{\t{Q}1} + \Delta_{\t{Q}2}$ (dotted line and circle). The conditions of $\Delta_{\t{Q}3} \approx \Delta_{\t{Q}1} + \Delta_{\t{Q}2}$ happen at different flux values with and without coupling because of a re-normalization of the qubit energy levels due to the coupling.}
	\label{fig:anticrossing}
\end{figure}

To determine the strength of the three-body interaction mediated by the coupler based on circuit simulations, the following approach is used. Qubit flux biasing parameters are chosen to have $\bar{\Delta}_{\t{Q}1} = \bar{\Delta}_{\t{Q}2}$ and $\bar{\Delta}_{\t{Q}3} = \bar{\Delta}_{\t{Q}1} + \bar{\Delta}_{\t{Q}2}$. For this bias condition, the spectrum of the complete circuit Hamiltonian [Eq.~(1) in the main text] has an avoided-level crossing between the $4^\t{th}$ and $5^\t{th}$ energy levels, $\ket{\uparrow \uparrow \downarrow}$ and $\ket{\downarrow \downarrow \uparrow}$ where the $\uparrow$ or $\downarrow$ refers to the current state of one of the three qubits. The avoided level crossing arises due to the ZZZ interaction. This coupling results in a mutual repulsion between the levels.  Calculating the minimum spacing between these two avoided levels gives $2\, \abs{J_{123}}$. Figure \ref{fig:anticrossing} illustrates this procedure.

An alternative method to calculate the coupling relies on choosing biasing conditions for the qubits such that $\bar{\Delta}_{\t{Q}1} = \bar{\Delta}_{\t{Q}2} = \bar{\Delta}_{\t{Q}3}$ and $\bar{\epsilon}_{\t{Q}1} = \bar{\epsilon}_{\t{Q}2} = \bar{\epsilon}_{\t{Q}3} = 0$. If the coupler is biased such that it is mediating a pure three-body interaction, that is $\bar{\epsilon}'_{\t{Q}i}=0$ and $\bar{J}_{ij}=0$, the 8-lowest energy levels of the circuit will form two four-fold degenerate subspaces. These two subspacces correspond to even and odd parity of the qubit currents states, i.e. the even states are $\ket{\uparrow \uparrow \uparrow},~\ket{\uparrow \downarrow \downarrow},~\ket{\downarrow \downarrow \uparrow}$ and $\ket{\downarrow \uparrow \downarrow}$ and the odd states are $\ket{\downarrow \downarrow \downarrow},~\ket{\uparrow \uparrow \downarrow},~\ket{\downarrow \uparrow \uparrow}$ and $\ket{\uparrow \downarrow \uparrow}$. These two subspaces are coupled by the ZZZ operator. Calculating the energy separation between these two subspaces gives $2\, \abs{J_{123}}$. This method was used to extract the coupling strength in Fig. 2 in the main text. 

\section{Spin Model}
\subsection{Correspondence Between Circuit Parameters and Spin Parameters}
\begin{figure}[h]
	\centering
    \includegraphics[width=4cm]{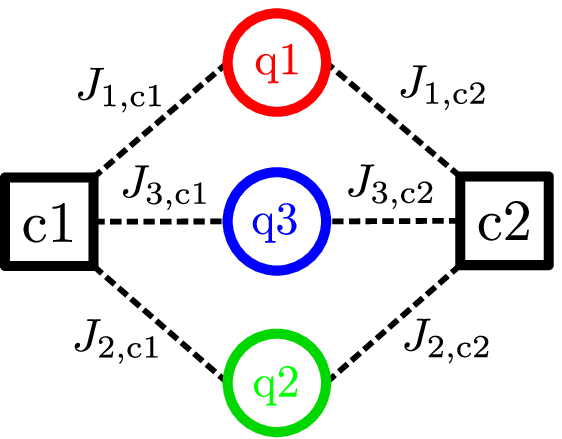}
    \caption{Spin-model schematic of the coupler plus qubits system. The circles and squares represent the qubit and coupler spins, respectively. The dashed lines represent bare two-body interactions.}
    \label{fig:CouplerSpin}
\end{figure}

The spin model for the coupled system presented in the main text is schematicized in Fig.~\ref{fig:CouplerSpin}. To obtain the correspondence between the circuit parameters and the spin parameters, we numerically diagonalize the circuit Hamiltonian of each isolated qubit or coupler at each bias point. Then, we take the two lowest energy states as the spin subspace of each device. 

The excitation energy at each bias point is defined as the difference between the ground state energy and the first excited state energy, written as 
\be
E^{01}_{\alpha\,i}\left(f_{\alpha\,i}^{\text{m}},\, f_{\alpha\,i}^{\text{s}} \right) = E^{1}_{\alpha\,i}\left(f_{\alpha\,i}^{\text{m}},\, f_{\alpha\,i}^{\text{s}} \right) - E^{0}_{\alpha\,i}\left(f_{\alpha\,i}^{\text{m}},\, f_{\alpha\,i}^{\text{s}} \right).
\ee
For any qubit or coupler, the energy bias spin parameter $\bar{\epsilon}_{\alpha\,i}$, is taken to be
\be
\bar{\epsilon}_{\alpha\,i} = \sqrt{E^{01}_{\alpha\,i}\left(f_{\alpha\,i}^{\text{m}},\,f_{\alpha\,i}^{\text{s}}\right)^2 - \bar{\Delta}_{\alpha\,i}\left(f_{\alpha\,i}^{\text{m}},\, f_{\alpha\,i}^{\text{s}} \right)^2}.
\ee
For any qubit or coupler, the transverse field spin parameter $\bar{\Delta}_{\alpha\,i}$ is taken to be 
\be
\bar{\Delta}_{\alpha\,i} = E^{01}_{\alpha\,i}\left(f_{\alpha\,i}^{\text{s}},\,0.5\right).
\ee
For any qubit - coupler pair, the direct 2-body interaction strength spin parameter $\bar{J}_{\t{Q}i,\t{C}j}$ is
\be
\bar{J}_{\t{Q}i,\t{C}j} = M_{i\,j}\,I^\text{p,m}_{\t{q}\,i} \l( f_{\t{q}\,i}^{\text{m}},\,f_{\t{q}\,i}^{\text{s}} \r) I^\text{p,m}_{\t{c}\,j} \l( f_{\t{c}\,j}^{\text{m}},\,f_{\t{c}\,j}^{\text{s}} \r).
\ee 
and 
\be
\bar{J}_{\t{Q}3,\t{C}j} = \l( M_{3\,j}^\t{s1} + M_{3\,j}^\t{s2} \r)\,I^\text{p,m}_{\t{q}\,3} \l( f_{\t{q}\,i}^{\text{m}},\,f_{\t{q}\,i}^{\text{s}} \r) I^\text{p,s}_{\t{c}\,j} \l( f_{\t{c}\,j}^{\text{m}},\,f_{\t{c}\,j}^{\text{s}} \r).
\ee 
The persistent current in the devices, $I^{\text{p},\beta}_{\alpha\,i}$, is defined as 
\be
I^{\text{p},\beta}_{\alpha\,i} = \frac{1}{2}\, \l[ \abs{I_{\alpha\,i}^{0,\beta} \l(  f_{\alpha\,i}^{\t{m}},\, f_{\alpha\,i}^{\t{s}} \r)} + \abs{I_{\alpha\,i}^{1,\beta} \l( f_{\alpha\,i}^{\t{m}},\, f_{\alpha\,i}^{\t{s}} \r)} \r],
\ee
where $I_{\alpha\,i}^{n,\beta}$ for $n \in \l\{ \t{0, 1} \r\}$ are the eigenvalues of the operator representing the current in loop $\beta \in \l\{\t{m,s}\r\}$ $I^\beta_{\alpha\,i}$ of device $\alpha i$ in the energy basis of the ground and first excited states of the device. The operator representing the current in the secondary loop of the coupler is defined as $I^\t{s}_{\t{c}\,j} = \phi_0\,\l( \varphi_{\t{c}\,j}^\t{s2} / L_{\t{c}\,j}^\t{s2} - \varphi_{\t{c}\,j}^\t{s1} / L_{\t{c}\,j}^\t{s1} \r) / 2$.
\subsection{Effective Hamiltonian in the Spin Model}
We briefly outline the procedure used to obtain the effective qubit-qubit Hamiltonian shown in Eq.~(\ref{eq:HeffSpin}) in the main text. We consider the spin Hamiltonian in Eq.~(\ref{eq:HSpin}) of the main text
\be
\label{eq:Hspin}
\bar{H} = \bar{H}_\t{Q} + \bar{H}_\t{C} + \bar{H}_\t{Int},
\ee
We assume the interaction to be weak; specifically, we assume $\bra{0_\t{c}} \bar{H}_\t{int} \ket{0_\t{c}}\, / \omega_{0,1} \ll 1$, where $\omega_{0,1}$ is the coupler excitation energy and $\ket{0_\t{c}}$ represents the ground state of the coupler. 
The time evolution operator in the interaction picture is 
\be
\bar{U}(t,0) = T\, \t{exp}\l[ \frac{-\t{i}}{\hbar} \int_0^t \bar{H}_\t{int} \l( t' \r) \t{d} t' \r],
\ee
where $T$ specifies the time ordering.
The effective coupling between the qubits can be found by projecting the full time evolution operator onto the ground state of the coupler, $\ket{0_\t{c}}$.
The effective propagator is given by 
\ba
\label{eq:Ueff}
\bar{U}_\t{eff}(t,0) &\equiv \bra{0_\t{c}} \bar{U}( t, 0) \ket{0_\t{c}}\nn \\
&= T\, \t{exp} \l[ \frac{-\t{i}}{\hbar} \int_0^t \bar{H}_\t{eff} \l( t' \r) \t{d}t' \r].
\ea
The effective qubit Hamiltonian is found by expanding $\bar{U}_\t{eff}(t,0)$ perturbatively up to third order~\cite{S_Hutter_2006}.

Applying the procedure from Eqs.~(\ref{eq:Hspin}-\ref{eq:Ueff}) to the the spin model Hamiltonian in Eq.~(\ref{eq:HSpin}) of the main text, we obtain the effective Hamiltonian in the spin picture shown in Eq.~(\ref{eq:HeffSpin}) in the main text.

The first order correction term to the qubit bias is given by 
\be
\bar{\epsilon}'_{\t{Q}1} = \sum_{j}^2 \bar{J}_{\t{Q}1,\t{C}j} \bra{0_\t{c}} \sigma_{\t{C}\,j}^z \ket{0_\t{c}}.
\ee
A similar expression can be written for the biases of qubits 2 and 3.
The shift in the qubit tunneling at the second order is given by
\be
\bar{\Delta}'_{\t{Q}1} = - \frac{\bar{\Delta}_{\t{Q}1}\, \bar{J}_{1,\t{C}1}\, \bar{J}_{\t{Q}1,\t{C}2}}{2} \sum_{n} \frac{\bar{V}_{0,n}}{\omega_{0,n}^2},
\ee
where $\bar{V}_{0,n}$ is as defined in the main text. Similar expressions can be written for tunneling elements for qubits 2 and 3.

Two body couplings $\bar{J}_{1,3}$ and $\bar{J}_{2,3}$ between qubits 1 and 3, and 2 and 3 are given by
\be
\bar{J}_{13}= - \l( \bar{A}\, \bar{J}_{\t{Q}1,\t{C}1}\, \bar{J}_{\t{Q}3,\t{C}2} + \bar{A}'\, \bar{J}_{\t{Q}3,\t{C}1}\, \bar{J}_{\t{Q}1,\t{C}2} \r)
\ee
and
\be
\bar{J}_{23}= - \l( \bar{A}\, \bar{J}_{\t{Q}2,\t{C}1}\, \bar{J}_{\t{Q}3,\t{C}2} + \bar{A}'\, \bar{J}_{\t{Q}3,\t{C}1}\, \bar{J}_{\t{Q}2,\t{C}2} \r),
\ee
where 
\be
\bar{A} = \sum_{n} \frac{\bra{0_\t{c}} \sigma_{\t{C}\,1}^z \ket{n_\t{c}}\! \bra{n_\t{c}} \sigma_{\t{C}\,2}^x \ket{0_\t{c}}}{\omega_{0,n}},
\ee
\be
\bar{A}' = \sum_{n} \frac{ \bra{0_\t{c}} \sigma_{\t{C}\,1}^x \ket{n_\t{c}}\! \bra{n_\t{c}} \sigma_{\t{C}\,2}^z \ket{0_\t{c}}}{\omega_{0,n}},
\ee
The conditions for cancellation of the two-body interactions between qubits 1 and 3, and 2 and 3 are
$\bar{A}\, \bar{J}_{\t{Q}1,\t{C}1}\, \bar{J}_{\t{Q}3,\t{C}2} = - \bar{A}'\, \bar{J}_{\t{Q}3,\t{C}1}\, \bar{J}_{\t{Q}1,\t{C}2}$ and $\bar{A}\, \bar{J}_{\t{Q}2,\t{C}1}\, \bar{J}_{\t{Q}3,\t{C}2} = - \bar{A}'\, \bar{J}_{\t{Q}3,\t{C}1}\, \bar{J}_{\t{Q}2,\t{C}2}$.
The effective three-body coupling $\bar{J}_{123}$ arises at the third order in perturbation theory. The expression for the effective coupling in the spin picture can be written as
\ba
\label{eq:JBar123}
\bar{J}_{123}  &= \sum_{i=1}^2 \sum_{j=1}^2 \sum_{k=1}^2\, \bar{J}_{\t{Q}1,\t{C}i}\,  \bar{J}_{\t{Q}2,\t{C}j}\,  \bar{J}_{\t{Q}3,\t{C}k}\,\nn\\
& \quad \times \sum_n \sum_m \frac{\bra{0_\t{c}} \sigma_{\t{C}\,i}^z \ket{n_\t{c}}\! \bra{n_\t{c}} \sigma_{\t{C}\,j}^z \ket{m_\t{c}}\! \bra{m_\t{c}} \sigma_{\t{C}\,k}^x \ket{0_\t{c}}}{\omega_{0,n}\, \omega_{0,m}}
\ea
where $\ket{m_\t{c}}$ and $\ket{n_\t{c}}$ are energy eigenstate of the coupler. 

\section{Sensitivity of the Coupler to Noise}

We determine the variations in the coupling and bias parameters induced by flux noise in the coupler loops. The Gaussian simulated flux noise environment for loop $\beta$ in device $\alpha\,i$ induces fluctuations in bias flux with a normal distribution of variance given by
\be
\delta \Phi_{\alpha\,i}^\beta = \int_{\omega_\t{lo}}^{\omega_\t{hi}} \t{d} \omega\, \frac{A_{\alpha\,i}^\beta}{\omega^{0.9}},
\ee
where $\omega_\t{hi} = 10$~GHz is the high-frequency cut-off, $\omega_\t{lo} = 2\pi/T_{\text{f}}$ is the low-frequency cut-off, where $T_{\t{f}}$ is the total experiment time which we have taken to be 1 $\mu$s, $A_{\alpha\,i}^\beta$ is an experimental flux noise amplitude of loop $\beta$ of device $\alpha\,i$. We have determined the value of the $A_{\alpha\,i}^\beta$, reported in the main text, by measuring the length-over-width ratio of the superconducting wires making up realistic computer assisted designs of the devices, and scaling this ratio with respect to established experimental values \cite{S_yan_2016_fluxqubitrevisited,S_weber_2017_coherentcoupledqubits,S_bialczak_2007_fluxnoisejosephson}. We model the spread in coupling and bias parameters when the system is subjected to this realistic noisy environment. The standard deviation of the parameter due to noise on the 4 loops of the coupling circuit and the qubit loops are reported in table \ref{tab:paramSpread}.

\begin{table}[h!]
\caption{Annealing parameter standard deviations due to uncorrelated flux noise in all coupler loops.}
\label{tab:paramSpread}
\centering
\begin{tabular}{ C{2.5cm} C{2.5cm} C{2.5cm} } 
 \hline\hline
 Parameter & Nominal value (GHz) & Standard deviation (GHz) \\
 \hline
 $\tilde{h}1$ & 0.000 & 0.101 \\ 
 $\tilde{h}2$ & 0.000 & 0.102  \\
 $\tilde{h}3$ & 0.000 & 0.006  \\
 $\tilde{J}12$ & 0.000 & 0.012 \\
 $\tilde{J}13$ & 0.000 & 0.015 \\
 $\tilde{J}23$ & 0.000 & 0.015 \\
 $\tilde{J}123$ & 0.915 & 0.003 \\
 \hline\hline
\end{tabular}
\end{table}

\section{Robustness of the Coupler to fabrication variations}
We determine the effect of fabrication variations on the coupler. We specifically observe the effect on the coupling strength and on the ability to cancel the two-body interactions when variations are introduced in the mutual inductances between the coupler and the qubits, the self inductances of the coupler and the Josephson junctions in the coupler circuit. 

The mutual inductances between the qubits and SQUIDs and the self-inductances of the SQUIDs were sampled randomly from a normal distribution with a 95\% confidence interval within 3\% of their nominal value. The coupling strengths $J_{ij}$ and $J_{123}$ were calculated with the Born-Oppenheimer method at the same bias point ($f_{\t{c}\,1}^{\t{m}}=f_{\t{c}\,2}^{\t{m}}=0.5$ and $f_{\t{c}\,1}^{\t{s}}=-f_{\t{c}\,2}^{\t{s}}=0.3$) for all random instances. This was repeated for 100 random instances.

All the Josephson junction critical currents in the rf-SQUIDs were sampled randomly from a normal distribution with a 95\% confidence interval within 3\% of their nominal value. For each random instance, the coupling strengths $J_{ij}$ and $J_{123}$ were calculated with the Born-Oppenheimer method at the same bias point ($f_{\t{c}\,1}^{\t{m}}=f_{\t{c}\,2}^{\t{m}}=0.5$ and $f_{\t{c}\,1}^{\t{s}}=-f_{\t{c}\,2}^{\t{s}}=0.3$). The asymmetry in the junctions affects the cancellation of the two-body interactions. New bias conditions where all two-body interactions are suppressed to within 10 MHz and where the three-body interaction is tunable were found by local optimization around the nominal bias point.

\section{Annealing simulations}
The full circuit annealing simulation in the main text is done in the following way. We start by diagonalizing the individual qubits to determine the relation between $f_{\t{q}\,i}^\t{s}$ and $f_{\t{c}\,i}^\t{s}$, and $\l(1 - s\r)\, \Delta_i$ in the annealing Hamiltonian. To account for the effect of the coupler on the qubits, we re-normalize the qubit inductance with a term proportional to the coupler susceptibility, as is discussed in Refs.~\cite{S_vandenbrink_2005_mediatedtunablecoupling,S_harris_2009_compoundjosephsonjunctioncoupler,S_weber_2017_coherentcoupledqubits}. This re-normalization is
\be
L_{\t{q}\,i}^\t{m} \rightarrow L_{\t{q}\,i}^\t{m} - \sum_{j=1}^2 M_{ij}^2\, \frac{\partial \bra{0_\t{c}} I_{\t{q}i}^\t{m} \ket{0_\t{c}}}{\partial f_{\t{c}\,j}^\t{m}},
\ee
for $i \in \l\{ 1,2 \r\}$, and
\be
L_{\t{q}\,3}^\t{m} \rightarrow L_{\t{q}\,3}^\t{m} - \sum_{j=1}^2 M_{3j}^2\, \frac{\partial \bra{0_\t{c}} I_{\t{q}3}^\t{m} \ket{0_\t{c}}}{\partial f_{\t{c}\,j}^\t{s}}.
\ee
We next determine the persistent current of these re-normalized qubits. These persistent currents are then used to determine the relation between $f_{\t{q}\,i}^\t{s}$ and $f_{\t{c}\,i}^\t{s}$, and $s\, J_{123}$ in the annealing Hamiltonian. We use the Born-Oppenheimer method described in Eq.~(6) in the main text, with the qubit persistent currents just calculated. This procedure is then repeated for all values of $s$. 




\begin{thebibliography}{37}%
\makeatletter
\providecommand \@ifxundefined [1]{%
 \@ifx{#1\undefined}
}%
\providecommand \@ifnum [1]{%
 \ifnum #1\expandafter \@firstoftwo
 \else \expandafter \@secondoftwo
 \fi
}%
\providecommand \@ifx [1]{%
 \ifx #1\expandafter \@firstoftwo
 \else \expandafter \@secondoftwo
 \fi
}%
\providecommand \natexlab [1]{#1}%
\providecommand \enquote  [1]{``#1''}%
\providecommand \bibnamefont  [1]{#1}%
\providecommand \bibfnamefont [1]{#1}%
\providecommand \citenamefont [1]{#1}%
\providecommand \href@noop [0]{\@secondoftwo}%
\providecommand \href [0]{\begingroup \@sanitize@url \@href}%
\providecommand \@href[1]{\@@startlink{#1}\@@href}%
\providecommand \@@href[1]{\endgroup#1\@@endlink}%
\providecommand \@sanitize@url [0]{\catcode `\\12\catcode `\$12\catcode
  `\&12\catcode `\#12\catcode `\^12\catcode `\_12\catcode `\%12\relax}%
\providecommand \@@startlink[1]{}%
\providecommand \@@endlink[0]{}%
\providecommand \url  [0]{\begingroup\@sanitize@url \@url }%
\providecommand \@url [1]{\endgroup\@href {#1}{\urlprefix }}%
\providecommand \urlprefix  [0]{URL }%
\providecommand \Eprint [0]{\href }%
\providecommand \doibase [0]{http://dx.doi.org/}%
\providecommand \selectlanguage [0]{\@gobble}%
\providecommand \bibinfo  [0]{\@secondoftwo}%
\providecommand \bibfield  [0]{\@secondoftwo}%
\providecommand \translation [1]{[#1]}%
\providecommand \BibitemOpen [0]{}%
\providecommand \bibitemStop [0]{}%
\providecommand \bibitemNoStop [0]{.\EOS\space}%
\providecommand \EOS [0]{\spacefactor3000\relax}%
\providecommand \BibitemShut  [1]{\csname bibitem#1\endcsname}%
\let\auto@bib@innerbib\@empty
\bibitem [{\citenamefont {Navr\'atil}\ and\ \citenamefont
  {Ormand}(2003)}]{Navr2003}%
  \BibitemOpen
  \bibfield  {author} {\bibinfo {author} {\bibfnamefont {P.}~\bibnamefont
  {Navr\'atil}}\ and\ \bibinfo {author} {\bibfnamefont {W.~E.}\ \bibnamefont
  {Ormand}},\ }\href {\doibase 10.1103/PhysRevC.68.034305} {\bibfield
  {journal} {\bibinfo  {journal} {Phys. Rev. C}\ }\textbf {\bibinfo {volume}
  {68}},\ \bibinfo {pages} {034305} (\bibinfo {year} {2003})}\BibitemShut
  {NoStop}%
\bibitem [{\citenamefont {von Stecher}(2011)}]{EfimovTrimer2011}%
  \BibitemOpen
  \bibfield  {author} {\bibinfo {author} {\bibfnamefont {J.}~\bibnamefont {von
  Stecher}},\ }\href {\doibase 10.1103/PhysRevLett.107.200402} {\bibfield
  {journal} {\bibinfo  {journal} {Phys. Rev. Lett.}\ }\textbf {\bibinfo
  {volume} {107}},\ \bibinfo {pages} {200402} (\bibinfo {year}
  {2011})}\BibitemShut {NoStop}%
\bibitem [{\citenamefont {Kitaev}(2003)}]{KITAEV20032}%
  \BibitemOpen
  \bibfield  {author} {\bibinfo {author} {\bibfnamefont {A.}~\bibnamefont
  {Kitaev}},\ }\href@noop {} {\bibfield  {journal} {\bibinfo  {journal} {Ann.
  Phys.}\ }\textbf {\bibinfo {volume} {303}},\ \bibinfo {pages} {2 } (\bibinfo
  {year} {2003})}\BibitemShut {NoStop}%
\bibitem [{\citenamefont {Jiang}\ and\ \citenamefont
  {Rieffel}(2017)}]{jiang_2017_noncommutingtwolocalhamiltonians}%
  \BibitemOpen
  \bibfield  {author} {\bibinfo {author} {\bibfnamefont {Z.}~\bibnamefont
  {Jiang}}\ and\ \bibinfo {author} {\bibfnamefont {E.~G.}\ \bibnamefont
  {Rieffel}},\ }\href {\doibase 10.1007/s11128-017-1527-9} {\bibfield
  {journal} {\bibinfo  {journal} {Quantum Inf. Process.}\ }\textbf {\bibinfo
  {volume} {16}},\ \bibinfo {pages} {89} (\bibinfo {year} {2017})}\BibitemShut
  {NoStop}%
\bibitem [{\citenamefont {Kempe}\ \emph {et~al.}(2006)\citenamefont {Kempe},
  \citenamefont {Kitaev},\ and\ \citenamefont {Regev}}]{Kempe2005}%
  \BibitemOpen
  \bibfield  {author} {\bibinfo {author} {\bibfnamefont {J.}~\bibnamefont
  {Kempe}}, \bibinfo {author} {\bibfnamefont {A.}~\bibnamefont {Kitaev}}, \
  and\ \bibinfo {author} {\bibfnamefont {O.}~\bibnamefont {Regev}},\ }\href
  {\doibase 10.1137/S0097539704445226} {\bibfield  {journal} {\bibinfo
  {journal} {SIAM J. Comput.}\ }\textbf {\bibinfo {volume} {35}},\ \bibinfo
  {pages} {1070} (\bibinfo {year} {2006})}\BibitemShut {NoStop}%
\bibitem [{\citenamefont {Linden}\ \emph {et~al.}(2010)\citenamefont {Linden},
  \citenamefont {Popescu},\ and\ \citenamefont {Skrzypczyk}}]{Linden2010}%
  \BibitemOpen
  \bibfield  {author} {\bibinfo {author} {\bibfnamefont {N.}~\bibnamefont
  {Linden}}, \bibinfo {author} {\bibfnamefont {S.}~\bibnamefont {Popescu}}, \
  and\ \bibinfo {author} {\bibfnamefont {P.}~\bibnamefont {Skrzypczyk}},\
  }\href {\doibase 10.1103/PhysRevLett.105.130401} {\bibfield  {journal}
  {\bibinfo  {journal} {Phys. Rev. Lett.}\ }\textbf {\bibinfo {volume} {105}},\
  \bibinfo {pages} {130401} (\bibinfo {year} {2010})}\BibitemShut {NoStop}%
\bibitem [{\citenamefont {Babbush}\ \emph {et~al.}(2014)\citenamefont
  {Babbush}, \citenamefont {Love},\ and\ \citenamefont
  {Aspuru-Guzik}}]{AlanA2014}%
  \BibitemOpen
  \bibfield  {author} {\bibinfo {author} {\bibfnamefont {R.}~\bibnamefont
  {Babbush}}, \bibinfo {author} {\bibfnamefont {P.}~\bibnamefont {Love}}, \
  and\ \bibinfo {author} {\bibfnamefont {A.}~\bibnamefont {Aspuru-Guzik}},\
  }\href {\doibase 10.1038/srep06603} {\bibfield  {journal} {\bibinfo
  {journal} {Sci. Rep.}\ }\textbf {\bibinfo {volume} {4}},\ \bibinfo {pages}
  {6603 EP} (\bibinfo {year} {2014})}\BibitemShut {NoStop}%
\bibitem [{\citenamefont {Peng}\ \emph {et~al.}(2009)\citenamefont {Peng} \emph
  {et~al.}}]{Quantsim2009}%
  \BibitemOpen
  \bibfield  {author} {\bibinfo {author} {\bibfnamefont {X.}~\bibnamefont
  {Peng}} \emph {et~al.},\ }\href {\doibase 10.1103/PhysRevLett.103.140501}
  {\bibfield  {journal} {\bibinfo  {journal} {Phys. Rev. Lett.}\ }\textbf
  {\bibinfo {volume} {103}},\ \bibinfo {pages} {140501} (\bibinfo {year}
  {2009})}\BibitemShut {NoStop}%
\bibitem [{\citenamefont {Porras}\ and\ \citenamefont
  {I~Cirac}(2004)}]{Porras2004}%
  \BibitemOpen
  \bibfield  {author} {\bibinfo {author} {\bibfnamefont {D.}~\bibnamefont
  {Porras}}\ and\ \bibinfo {author} {\bibfnamefont {J.}~\bibnamefont
  {I~Cirac}},\ }\href {\doibase 10.1103/PhysRevLett.92.207901} {\bibfield
  {journal} {\bibinfo  {journal} {Phys. Rev. Lett.}\ }\textbf {\bibinfo
  {volume} {92}},\ \bibinfo {pages} {207901} (\bibinfo {year}
  {2004})}\BibitemShut {NoStop}%
\bibitem [{\citenamefont {Bermudez}\ \emph {et~al.}(2009)\citenamefont
  {Bermudez}, \citenamefont {Porras},\ and\ \citenamefont
  {Martin-Delgado}}]{Bermudez2009}%
  \BibitemOpen
  \bibfield  {author} {\bibinfo {author} {\bibfnamefont {A.}~\bibnamefont
  {Bermudez}}, \bibinfo {author} {\bibfnamefont {D.}~\bibnamefont {Porras}}, \
  and\ \bibinfo {author} {\bibfnamefont {M.~A.}\ \bibnamefont
  {Martin-Delgado}},\ }\href {\doibase 10.1103/PhysRevA.79.060303} {\bibfield
  {journal} {\bibinfo  {journal} {Phys. Rev. A}\ }\textbf {\bibinfo {volume}
  {79}},\ \bibinfo {pages} {060303} (\bibinfo {year} {2009})}\BibitemShut
  {NoStop}%
\bibitem [{\citenamefont {Pachos}\ and\ \citenamefont
  {Rico}(2004)}]{Pachos2004}%
  \BibitemOpen
  \bibfield  {author} {\bibinfo {author} {\bibfnamefont {J.~K.}\ \bibnamefont
  {Pachos}}\ and\ \bibinfo {author} {\bibfnamefont {E.}~\bibnamefont {Rico}},\
  }\href {\doibase 10.1103/PhysRevA.70.053620} {\bibfield  {journal} {\bibinfo
  {journal} {Phys. Rev. A}\ }\textbf {\bibinfo {volume} {70}},\ \bibinfo
  {pages} {053620} (\bibinfo {year} {2004})}\BibitemShut {NoStop}%
\bibitem [{\citenamefont {Semi{\~a}o}\ and\ \citenamefont
  {Paternostro}(2012)}]{Semião2012}%
  \BibitemOpen
  \bibfield  {author} {\bibinfo {author} {\bibfnamefont {F.~L.}\ \bibnamefont
  {Semi{\~a}o}}\ and\ \bibinfo {author} {\bibfnamefont {M.}~\bibnamefont
  {Paternostro}},\ }\href {\doibase 10.1007/s11128-011-0232-3} {\bibfield
  {journal} {\bibinfo  {journal} {Quantum Inf. Process.}\ }\textbf {\bibinfo
  {volume} {11}},\ \bibinfo {pages} {67} (\bibinfo {year} {2012})}\BibitemShut
  {NoStop}%
\bibitem [{\citenamefont {B\"uchler}\ \emph {et~al.}(2007)\citenamefont
  {B\"uchler}, \citenamefont {Micheli},\ and\ \citenamefont
  {Zoller}}]{buchler_2007_threebodyinteractionscold}%
  \BibitemOpen
  \bibfield  {author} {\bibinfo {author} {\bibfnamefont {H.~P.}\ \bibnamefont
  {B\"uchler}}, \bibinfo {author} {\bibfnamefont {A.}~\bibnamefont {Micheli}},
  \ and\ \bibinfo {author} {\bibfnamefont {P.}~\bibnamefont {Zoller}},\ }\href
  {\doibase 10.1038/nphys678} {\bibfield  {journal} {\bibinfo  {journal} {Nat.
  Phys.}\ }\textbf {\bibinfo {volume} {3}},\ \bibinfo {pages} {726} (\bibinfo
  {year} {2007})}\BibitemShut {NoStop}%
\bibitem [{\citenamefont {Chen}\ and\ \citenamefont {Li}(2012)}]{Chen_2012}%
  \BibitemOpen
  \bibfield  {author} {\bibinfo {author} {\bibfnamefont {Y.-X.}\ \bibnamefont
  {Chen}}\ and\ \bibinfo {author} {\bibfnamefont {S.-W.}\ \bibnamefont {Li}},\
  }\href {\doibase 10.1209/0295-5075/97/40003} {\bibfield  {journal} {\bibinfo
  {journal} {Europhys. Lett.}\ }\textbf {\bibinfo {volume} {97}},\ \bibinfo
  {pages} {40003} (\bibinfo {year} {2012})}\BibitemShut {NoStop}%
\bibitem [{\citenamefont {Kafri}\ \emph {et~al.}(2017)\citenamefont {Kafri}
  \emph {et~al.}}]{kafri_2017_tunableinductivecoupling}%
  \BibitemOpen
  \bibfield  {author} {\bibinfo {author} {\bibfnamefont {D.}~\bibnamefont
  {Kafri}} \emph {et~al.},\ }\href {\doibase 10.1103/PhysRevA.95.052333}
  {\bibfield  {journal} {\bibinfo  {journal} {Phys. Rev. A}\ }\textbf {\bibinfo
  {volume} {95}},\ \bibinfo {pages} {052333} (\bibinfo {year}
  {2017})}\BibitemShut {NoStop}%
\bibitem [{\citenamefont {Cho}\ and\ \citenamefont
  {Kim}(2008)}]{cho_2008_macroscopicmanyqubitinteractions}%
  \BibitemOpen
  \bibfield  {author} {\bibinfo {author} {\bibfnamefont {S.~Y.}\ \bibnamefont
  {Cho}}\ and\ \bibinfo {author} {\bibfnamefont {M.~D.}\ \bibnamefont {Kim}},\
  }\href {\doibase 10.1103/PhysRevB.77.212506} {\bibfield  {journal} {\bibinfo
  {journal} {Phys. Rev. B}\ }\textbf {\bibinfo {volume} {77}},\ \bibinfo
  {pages} {212506} (\bibinfo {year} {2008})}\BibitemShut {NoStop}%
\bibitem [{\citenamefont
  {Sameti}(2017)}]{sametiSuperconductingQuantumSimulator2017}%
  \BibitemOpen
  \bibfield  {author} {\bibinfo {author} {\bibfnamefont {M.~o.}\ \bibnamefont
  {Sameti}},\ }\href {\doibase 10.1103/PhysRevA.95.042330} {\bibfield
  {journal} {\bibinfo  {journal} {Phys. Rev. A}\ }\textbf {\bibinfo {volume}
  {95}},\ \bibinfo {pages} {042330} (\bibinfo {year} {2017})}\BibitemShut
  {NoStop}%
\bibitem [{\citenamefont {Puri}\ \emph {et~al.}(2017)\citenamefont {Puri} \emph
  {et~al.}}]{puri_2017_quantumannealingalltoalla}%
  \BibitemOpen
  \bibfield  {author} {\bibinfo {author} {\bibfnamefont {S.}~\bibnamefont
  {Puri}} \emph {et~al.},\ }\href {\doibase 10.1038/ncomms15785} {\bibfield
  {journal} {\bibinfo  {journal} {Nat. Commun.}\ }\textbf {\bibinfo {volume}
  {8}},\ \bibinfo {pages} {ncomms15785} (\bibinfo {year} {2017})}\BibitemShut
  {NoStop}%
\bibitem [{\citenamefont {Strand}\ \emph {et~al.}(2017)\citenamefont {Strand}
  \emph {et~al.}}]{ngcZZZAPS}%
  \BibitemOpen
  \bibfield  {author} {\bibinfo {author} {\bibfnamefont {J.}~\bibnamefont
  {Strand}} \emph {et~al.},\ }in\ \href@noop {} {\emph {\bibinfo {booktitle}
  {American Physical Society, March Meeting, B51.00009}}}\ (\bibinfo {year}
  {2017})\BibitemShut {NoStop}%
\bibitem [{\citenamefont {Sch\"ondorf}\ and\ \citenamefont
  {Wilhelm}(2018)}]{schondorf_2018_fourlocalinteractionssuperconducting}%
  \BibitemOpen
  \bibfield  {author} {\bibinfo {author} {\bibfnamefont {M.}~\bibnamefont
  {Sch\"ondorf}}\ and\ \bibinfo {author} {\bibfnamefont {F.~K.}\ \bibnamefont
  {Wilhelm}},\ }\href@noop {} {\bibfield  {journal} {\bibinfo  {journal}
  {arXiv:1811.07683}\ } (\bibinfo {year} {2018})}\BibitemShut {NoStop}%
\bibitem [{\citenamefont {Kerman}(2018)}]{MITLL_ZZZZ_APS}%
  \BibitemOpen
  \bibfield  {author} {\bibinfo {author} {\bibfnamefont {A.~J.}\ \bibnamefont
  {Kerman}},\ }in\ \href@noop {} {\emph {\bibinfo {booktitle} {American
  Physical Society, March Meeting, C26.00001}}}\ (\bibinfo {year}
  {2018})\BibitemShut {NoStop}%
\bibitem [{\citenamefont {Chancellor}\ \emph {et~al.}(2017)\citenamefont
  {Chancellor}, \citenamefont {Zohren},\ and\ \citenamefont
  {Warburton}}]{chancellor_2017_circuitdesignmultibody}%
  \BibitemOpen
  \bibfield  {author} {\bibinfo {author} {\bibfnamefont {N.}~\bibnamefont
  {Chancellor}}, \bibinfo {author} {\bibfnamefont {S.}~\bibnamefont {Zohren}},
  \ and\ \bibinfo {author} {\bibfnamefont {P.~A.}\ \bibnamefont {Warburton}},\
  }\href {\doibase 10.1038/s41534-017-0022-6} {\bibfield  {journal} {\bibinfo
  {journal} {npj Quantum Inf.}\ }\textbf {\bibinfo {volume} {3}},\ \bibinfo
  {pages} {21} (\bibinfo {year} {2017})}\BibitemShut {NoStop}%
\bibitem [{\citenamefont {Harris}\ \emph {et~al.}(2010)\citenamefont {Harris}
  \emph {et~al.}}]{harris_2010_experimentaldemonstrationrobust}%
  \BibitemOpen
  \bibfield  {author} {\bibinfo {author} {\bibfnamefont {R.}~\bibnamefont
  {Harris}} \emph {et~al.},\ }\href {\doibase 10.1103/PhysRevB.81.134510}
  {\bibfield  {journal} {\bibinfo  {journal} {Phys. Rev. B}\ }\textbf {\bibinfo
  {volume} {81}},\ \bibinfo {pages} {134510} (\bibinfo {year} {2010})},\
  \bibinfo {note} {00080}\BibitemShut {NoStop}%
\bibitem [{\citenamefont {Weber}\ \emph {et~al.}(2017)\citenamefont {Weber}
  \emph {et~al.}}]{weber_2017_coherentcoupledqubits}%
  \BibitemOpen
  \bibfield  {author} {\bibinfo {author} {\bibfnamefont {S.~J.}\ \bibnamefont
  {Weber}} \emph {et~al.},\ }\href@noop {} {\bibfield  {journal} {\bibinfo
  {journal} {Phys. Rev. Appl.}\ }\textbf {\bibinfo {volume} {8}},\ \bibinfo
  {pages} {014004} (\bibinfo {year} {2017})}\BibitemShut {NoStop}%
\bibitem [{\citenamefont {Yan}\ \emph {et~al.}(2016)\citenamefont {Yan} \emph
  {et~al.}}]{yan_2016_fluxqubitrevisited}%
  \BibitemOpen
  \bibfield  {author} {\bibinfo {author} {\bibfnamefont {F.}~\bibnamefont
  {Yan}} \emph {et~al.},\ }\href {\doibase 10.1038/ncomms12964} {\bibfield
  {journal} {\bibinfo  {journal} {Nat. Commun.}\ }\textbf {\bibinfo {volume}
  {7}},\ \bibinfo {pages} {ncomms12964} (\bibinfo {year} {2016})}\BibitemShut
  {NoStop}%
\bibitem [{\citenamefont {Poletto}\ \emph {et~al.}(2009)\citenamefont {Poletto}
  \emph {et~al.}}]{polettoTunableRfSQUID2009}%
  \BibitemOpen
  \bibfield  {author} {\bibinfo {author} {\bibfnamefont {S.}~\bibnamefont
  {Poletto}} \emph {et~al.},\ }\href {\doibase
  10.1088/0031-8949/2009/T137/014011} {\bibfield  {journal} {\bibinfo
  {journal} {Phys. Scr.}\ }\textbf {\bibinfo {volume} {T137}},\ \bibinfo
  {pages} {014011} (\bibinfo {year} {2009})}\BibitemShut {NoStop}%
\bibitem [{\citenamefont {Harris}\ \emph {et~al.}(2009)\citenamefont {Harris}
  \emph {et~al.}}]{harris_2009_compoundjosephsonjunctioncoupler}%
  \BibitemOpen
  \bibfield  {author} {\bibinfo {author} {\bibfnamefont {R.}~\bibnamefont
  {Harris}} \emph {et~al.},\ }\href {\doibase 10.1103/PhysRevB.80.052506}
  {\bibfield  {journal} {\bibinfo  {journal} {Phys. Rev. B}\ }\textbf {\bibinfo
  {volume} {80}},\ \bibinfo {pages} {052506} (\bibinfo {year}
  {2009})}\BibitemShut {NoStop}%
\bibitem [{\citenamefont {van~den Brink}\ \emph {et~al.}(2005)\citenamefont
  {van~den Brink}, \citenamefont {Berkley},\ and\ \citenamefont
  {Yalowsky}}]{vandenbrink_2005_mediatedtunablecoupling}%
  \BibitemOpen
  \bibfield  {author} {\bibinfo {author} {\bibfnamefont {A.~M.}\ \bibnamefont
  {van~den Brink}}, \bibinfo {author} {\bibfnamefont {A.~J.}\ \bibnamefont
  {Berkley}}, \ and\ \bibinfo {author} {\bibfnamefont {M.}~\bibnamefont
  {Yalowsky}},\ }\href {\doibase 10.1088/1367-2630/7/1/230} {\bibfield
  {journal} {\bibinfo  {journal} {New J. Phys.}\ }\textbf {\bibinfo {volume}
  {7}},\ \bibinfo {pages} {230} (\bibinfo {year} {2005})}\BibitemShut {NoStop}%
\bibitem [{\citenamefont {Harris}\ \emph {et~al.}(2007)\citenamefont {Harris}
  \emph {et~al.}}]{harris_2007_signmagnitudetunablecoupler}%
  \BibitemOpen
  \bibfield  {author} {\bibinfo {author} {\bibfnamefont {R.}~\bibnamefont
  {Harris}} \emph {et~al.},\ }\href {\doibase 10.1103/PhysRevLett.98.177001}
  {\bibfield  {journal} {\bibinfo  {journal} {Phys. Rev. Lett.}\ }\textbf
  {\bibinfo {volume} {98}},\ \bibinfo {pages} {177001} (\bibinfo {year}
  {2007})}\BibitemShut {NoStop}%
\bibitem [{\citenamefont {Massey}(1949)}]{Massey_1949}%
  \BibitemOpen
  \bibfield  {author} {\bibinfo {author} {\bibfnamefont {H.~S.~W.}\
  \bibnamefont {Massey}},\ }\href {\doibase 10.1088/0034-4885/12/1/311}
  {\bibfield  {journal} {\bibinfo  {journal} {Rep. Prog. Phys.}\ }\textbf
  {\bibinfo {volume} {12}},\ \bibinfo {pages} {248} (\bibinfo {year}
  {1949})}\BibitemShut {NoStop}%
\bibitem [{\citenamefont {Hutter}\ \emph {et~al.}(2006)\citenamefont {Hutter}
  \emph {et~al.}}]{Hutter_2006}%
  \BibitemOpen
  \bibfield  {author} {\bibinfo {author} {\bibfnamefont {C.}~\bibnamefont
  {Hutter}} \emph {et~al.},\ }\href {\doibase 10.1209/epl/i2006-10054-4}
  {\bibfield  {journal} {\bibinfo  {journal} {Europhys. Lett.}\ }\textbf
  {\bibinfo {volume} {74}},\ \bibinfo {pages} {1088} (\bibinfo {year}
  {2006})}\BibitemShut {NoStop}%
\bibitem [{\citenamefont {Kerman}()}]{JJSim}%
  \BibitemOpen
  \bibfield  {author} {\bibinfo {author} {\bibfnamefont {A.~J.}\ \bibnamefont
  {Kerman}},\ }\href@noop {} {}\bibinfo {note} {In preparation}\BibitemShut
  {NoStop}%
\bibitem [{\citenamefont {Yoshihara}\ \emph {et~al.}(2006)\citenamefont
  {Yoshihara} \emph {et~al.}}]{yoshihara_2006_decoherencefluxqubits}%
  \BibitemOpen
  \bibfield  {author} {\bibinfo {author} {\bibfnamefont {F.}~\bibnamefont
  {Yoshihara}} \emph {et~al.},\ }\href {\doibase 10.1103/PhysRevLett.97.167001}
  {\bibfield  {journal} {\bibinfo  {journal} {Phys. Rev.Lett.}\ }\textbf
  {\bibinfo {volume} {97}},\ \bibinfo {pages} {167001} (\bibinfo {year}
  {2006})}\BibitemShut {NoStop}%
\bibitem [{\citenamefont {Koch}\ \emph {et~al.}(2007)\citenamefont {Koch},
  \citenamefont {DiVincenzo},\ and\ \citenamefont
  {Clarke}}]{koch_2007_modelfluxnoise}%
  \BibitemOpen
  \bibfield  {author} {\bibinfo {author} {\bibfnamefont {R.~H.}\ \bibnamefont
  {Koch}}, \bibinfo {author} {\bibfnamefont {D.~P.}\ \bibnamefont
  {DiVincenzo}}, \ and\ \bibinfo {author} {\bibfnamefont {J.}~\bibnamefont
  {Clarke}},\ }\href {\doibase 10.1103/PhysRevLett.98.267003} {\bibfield
  {journal} {\bibinfo  {journal} {Phys. Rev. Lett.}\ }\textbf {\bibinfo
  {volume} {98}},\ \bibinfo {pages} {267003} (\bibinfo {year}
  {2007})}\BibitemShut {NoStop}%
\bibitem [{\citenamefont {Bialczak}\ \emph {et~al.}(2007)\citenamefont
  {Bialczak} \emph {et~al.}}]{bialczak_2007_fluxnoisejosephson}%
  \BibitemOpen
  \bibfield  {author} {\bibinfo {author} {\bibfnamefont {R.~C.}\ \bibnamefont
  {Bialczak}} \emph {et~al.},\ }\href {\doibase 10.1103/PhysRevLett.99.187006}
  {\bibfield  {journal} {\bibinfo  {journal} {Phys. Rev. Lett.}\ }\textbf
  {\bibinfo {volume} {99}},\ \bibinfo {pages} {187006} (\bibinfo {year}
  {2007})}\BibitemShut {NoStop}%
\bibitem [{\citenamefont {Perdomo-Ortiz}\ \emph {et~al.}(2016)\citenamefont
  {Perdomo-Ortiz} \emph
  {et~al.}}]{perdomo-ortiz_2016_determinationcorrectionpersistent}%
  \BibitemOpen
  \bibfield  {author} {\bibinfo {author} {\bibnamefont {Perdomo-Ortiz}} \emph
  {et~al.},\ }\href {\doibase 10.1038/srep18628} {\bibfield  {journal}
  {\bibinfo  {journal} {Sci. Rep.}\ }\textbf {\bibinfo {volume} {6}},\ \bibinfo
  {pages} {srep18628} (\bibinfo {year} {2016})}\BibitemShut {NoStop}%
\bibitem [{\citenamefont {Albash}\ and\ \citenamefont
  {Lidar}(2018)}]{albash_2018_adiabaticquantumcomputation}%
  \BibitemOpen
  \bibfield  {author} {\bibinfo {author} {\bibfnamefont {T.}~\bibnamefont
  {Albash}}\ and\ \bibinfo {author} {\bibfnamefont {D.~A.}\ \bibnamefont
  {Lidar}},\ }\href {\doibase 10.1103/RevModPhys.90.015002} {\bibfield
  {journal} {\bibinfo  {journal} {Rev. Mod. Phys.}\ }\textbf {\bibinfo {volume}
  {90}},\ \bibinfo {pages} {015002} (\bibinfo {year} {2018})}\BibitemShut
  {NoStop}%
\end{thebibliography}

\begin{thebibliography}{6}%
\makeatletter
\providecommand \@ifxundefined [1]{%
 \@ifx{#1\undefined}
}%
\providecommand \@ifnum [1]{%
 \ifnum #1\expandafter \@firstoftwo
 \else \expandafter \@secondoftwo
 \fi
}%
\providecommand \@ifx [1]{%
 \ifx #1\expandafter \@firstoftwo
 \else \expandafter \@secondoftwo
 \fi
}%
\providecommand \natexlab [1]{#1}%
\providecommand \enquote  [1]{``#1''}%
\providecommand \bibnamefont  [1]{#1}%
\providecommand \bibfnamefont [1]{#1}%
\providecommand \citenamefont [1]{#1}%
\providecommand \href@noop [0]{\@secondoftwo}%
\providecommand \href [0]{\begingroup \@sanitize@url \@href}%
\providecommand \@href[1]{\@@startlink{#1}\@@href}%
\providecommand \@@href[1]{\endgroup#1\@@endlink}%
\providecommand \@sanitize@url [0]{\catcode `\\12\catcode `\$12\catcode
  `\&12\catcode `\#12\catcode `\^12\catcode `\_12\catcode `\%12\relax}%
\providecommand \@@startlink[1]{}%
\providecommand \@@endlink[0]{}%
\providecommand \url  [0]{\begingroup\@sanitize@url \@url }%
\providecommand \@url [1]{\endgroup\@href {#1}{\urlprefix }}%
\providecommand \urlprefix  [0]{URL }%
\providecommand \Eprint [0]{\href }%
\providecommand \doibase [0]{http://dx.doi.org/}%
\providecommand \selectlanguage [0]{\@gobble}%
\providecommand \bibinfo  [0]{\@secondoftwo}%
\providecommand \bibfield  [0]{\@secondoftwo}%
\providecommand \translation [1]{[#1]}%
\providecommand \BibitemOpen [0]{}%
\providecommand \bibitemStop [0]{}%
\providecommand \bibitemNoStop [0]{.\EOS\space}%
\providecommand \EOS [0]{\spacefactor3000\relax}%
\providecommand \BibitemShut  [1]{\csname bibitem#1\endcsname}%
\let\auto@bib@innerbib\@empty
\bibitem [{\citenamefont {Hutter}\ \emph {et~al.}(2006)\citenamefont {Hutter}
  \emph {et~al.}}]{S_Hutter_2006}%
  \BibitemOpen
  \bibfield  {author} {\bibinfo {author} {\bibfnamefont {C.}~\bibnamefont
  {Hutter}} \emph {et~al.},\ }\href {\doibase 10.1209/epl/i2006-10054-4}
  {\bibfield  {journal} {\bibinfo  {journal} {Europhys. Lett.}\ }\textbf
  {\bibinfo {volume} {74}},\ \bibinfo {pages} {1088} (\bibinfo {year}
  {2006})}\BibitemShut {NoStop}%
\bibitem [{\citenamefont {Yan}\ \emph {et~al.}(2016)\citenamefont {Yan} \emph
  {et~al.}}]{S_yan_2016_fluxqubitrevisited}%
  \BibitemOpen
  \bibfield  {author} {\bibinfo {author} {\bibfnamefont {F.}~\bibnamefont
  {Yan}} \emph {et~al.},\ }\href {\doibase 10.1038/ncomms12964} {\bibfield
  {journal} {\bibinfo  {journal} {Nat. Commun.}\ }\textbf {\bibinfo {volume}
  {7}},\ \bibinfo {pages} {ncomms12964} (\bibinfo {year} {2016})}\BibitemShut
  {NoStop}%
\bibitem [{\citenamefont {Weber}\ \emph {et~al.}(2017)\citenamefont {Weber}
  \emph {et~al.}}]{S_weber_2017_coherentcoupledqubits}%
  \BibitemOpen
  \bibfield  {author} {\bibinfo {author} {\bibfnamefont {S.~J.}\ \bibnamefont
  {Weber}} \emph {et~al.},\ }\href@noop {} {\bibfield  {journal} {\bibinfo
  {journal} {Phys. Rev. Appl.}\ }\textbf {\bibinfo {volume} {8}},\ \bibinfo
  {pages} {014004} (\bibinfo {year} {2017})}\BibitemShut {NoStop}%
\bibitem [{\citenamefont {Bialczak}\ \emph {et~al.}(2007)\citenamefont
  {Bialczak} \emph {et~al.}}]{S_bialczak_2007_fluxnoisejosephson}%
  \BibitemOpen
  \bibfield  {author} {\bibinfo {author} {\bibfnamefont {R.~C.}\ \bibnamefont
  {Bialczak}} \emph {et~al.},\ }\href {\doibase 10.1103/PhysRevLett.99.187006}
  {\bibfield  {journal} {\bibinfo  {journal} {Phys. Rev. Lett.}\ }\textbf
  {\bibinfo {volume} {99}},\ \bibinfo {pages} {187006} (\bibinfo {year}
  {2007})}\BibitemShut {NoStop}%
\bibitem [{\citenamefont {van~den Brink}\ \emph {et~al.}(2005)\citenamefont
  {van~den Brink}, \citenamefont {Berkley},\ and\ \citenamefont
  {Yalowsky}}]{S_vandenbrink_2005_mediatedtunablecoupling}%
  \BibitemOpen
  \bibfield  {author} {\bibinfo {author} {\bibfnamefont {A.~M.}\ \bibnamefont
  {van~den Brink}}, \bibinfo {author} {\bibfnamefont {A.~J.}\ \bibnamefont
  {Berkley}}, \ and\ \bibinfo {author} {\bibfnamefont {M.}~\bibnamefont
  {Yalowsky}},\ }\href {\doibase 10.1088/1367-2630/7/1/230} {\bibfield
  {journal} {\bibinfo  {journal} {New J. Phys.}\ }\textbf {\bibinfo {volume}
  {7}},\ \bibinfo {pages} {230} (\bibinfo {year} {2005})}\BibitemShut {NoStop}%
\bibitem [{\citenamefont {Harris}\ \emph {et~al.}(2009)\citenamefont {Harris}
  \emph {et~al.}}]{S_harris_2009_compoundjosephsonjunctioncoupler}%
  \BibitemOpen
  \bibfield  {author} {\bibinfo {author} {\bibfnamefont {R.}~\bibnamefont
  {Harris}} \emph {et~al.},\ }\href {\doibase 10.1103/PhysRevB.80.052506}
  {\bibfield  {journal} {\bibinfo  {journal} {Phys. Rev. B}\ }\textbf {\bibinfo
  {volume} {80}},\ \bibinfo {pages} {052506} (\bibinfo {year}
  {2009})}\BibitemShut {NoStop}%
\end{thebibliography}

-----------
BIBLIOGRAPHY
-----------

\end{document}